\documentclass[twocolumn]{aastex62}
\usepackage{CJK}

\newcommand{\superc}{03fg-like}
\newcommand{\kms}{\,km\,s$^{-1}$}
\newcommand{\dm}{$\Delta m_{15} (B)$}
\newcommand{\sbv}{$s_{\mathrm{BV}}$}

\newcommand{\um}{$\mu$m}

\newcommand{\hei}{\ion{He}{1}}
\newcommand{\ci}{\ion{C}{1}}
\newcommand{\cii}{\ion{C}{2}}

\newcommand{\naid}{\ion{Na}{1}\,D}
\newcommand{\siii}{\ion{Si}{2}}

\newcommand{\feii}{\ion{Fe}{2}}
\newcommand{\feiii}{\ion{Fe}{3}}

\newcommand{\lam}{$\lambda$}
\newcommand{\nirad}{$^{56}$Ni}

\newcommand{\corad}{$^{56}$Co}

\newcommand{\sneia}{SNe\,Ia}
\newcommand{\snia}{SN\,Ia}
\newcommand{\msun}{$\mathrm{M}_{\odot}$}
\newcommand{\msunperyr}{$\mathrm{M}_{\odot}$\,yr$^{-1}$}

\newcommand{\mcore}{1.45}
\newcommand{\menv}{0.2}
\newcommand{\msi}{0.08}
\newcommand{\mnirad}{1.07}
\newcommand{\mdeflag}{0.2}

\shorttitle{LSQ14fmg: Slowest Rising SN\,I\MakeLowercase{a}}
\shortauthors{Hsiao et al.}

\begin{document}
\begin{CJK*}{UTF8}{bsmi}

\title{Carnegie Supernova Project II: The slowest rising Type Ia
  supernova LSQ14fmg\\and clues to the origin of
  super-Chandrasekhar/03fg-like events\footnote{This paper includes
    data gathered with the 1-m Swope and the 2.5-m du Pont telescopes
    at Las Campanas Observatory, Chile, and the Nordic Optical
    Telescope at the Observatorio del Roque de los Muchachos, La
    Palma, Spain}}

\author[0000-0003-1039-2928]{E. Y. Hsiao (蕭亦麒)}
\affil{Department of Physics, Florida State University, 77 Chieftan Way, Tallahassee, FL 32306, USA}

\author[0000-0002-4338-6586]{P. Hoeflich}
\affil{Department of Physics, Florida State University, 77 Chieftan Way, Tallahassee, FL 32306, USA}

\author[0000-0002-5221-7557]{C. Ashall}
\affil{Department of Physics, Florida State University, 77 Chieftan Way, Tallahassee, FL 32306, USA}

\author[0000-0002-3900-1452]{J. Lu}
\affil{Department of Physics, Florida State University, 77 Chieftan Way, Tallahassee, FL 32306, USA}

\author[0000-0001-6293-9062]{C. Contreras}
\affil{Carnegie Observatories, Las Campanas Observatory, Colina El Pino, Casilla 601, Chile}

\author[0000-0003-4625-6629]{C. R. Burns}
\affil{Observatories of the Carnegie Institution for Science, 813 Santa Barbara St, Pasadena, CA 91101, USA}

\author[0000-0003-2734-0796]{M. M. Phillips}
\affil{Carnegie Observatories, Las Campanas Observatory, Colina El Pino, Casilla 601, Chile}

\author[0000-0002-1296-6887]{L. Galbany}
\affil{Departamento de F\'isica Te\'orica y del Cosmos, Universidad de Granada, E-18071 Granada, Spain}

%alphabetical

\author[0000-0003-0227-3451]{J. P. Anderson}
\affil{European Southern Observatory, Alonso de C\'ordova 3107, Casilla 19, Santiago, Chile}

\author[0000-0003-0424-8719]{C. Baltay}
\affil{Physics Department, Yale University, 217 Prospect Street, New Haven, CT 06511, USA}

\author[0000-0001-5393-1608]{E. Baron}
\affil{Homer L. Dodge Department of Physics and Astronomy, 440 West Brooks Street, Room 100, Norman, OK 73019, USA}

\author{S. Castell\'{o}n}
\affil{Carnegie Observatories, Las Campanas Observatory, Colina El Pino, Casilla 601, Chile}

\author[0000-0002-2806-5821]{S. Davis}
\affil{Department of Physics, Florida State University, 77 Chieftan Way, Tallahassee, FL 32306, USA}

\author[0000-0003-3431-9135]{Wendy L. Freedman}
\affil{Department of Astronomy and Astrophysics, University of Chicago, 5640 S. Ellis Avenue, Chicago, IL 60637, USA}

\author[0000-0002-8526-3963]{C. Gall}
\affil{DARK, Niels Bohr Institute, University of Copenhagen, Lyngbyvej 2, 2100, Copenhagen \O, Denmark}

\author{C. Gonzalez}
\affil{Carnegie Observatories, Las Campanas Observatory, Colina El Pino, Casilla 601, Chile}

\author[0000-0002-9154-3136]{M. L. Graham}
\affil{DiRAC Institute, Department of Astronomy, University of Washington, Box 351580, U.W., Seattle, WA 98195, USA}

\author[0000-0001-7981-8320]{M. Hamuy}
\affil{Universidad de Chile, Departamento de Astronom\'{i}a, Casilla 36-D, Santiago, Chile}

\author[0000-0001-9206-3460]{T. W.-S. Holoien}
\affil{Observatories of the Carnegie Institution for Science, 813 Santa Barbara St, Pasadena, CA 91101, USA}

\author[0000-0001-6209-838X]{E. Karamehmetoglu}
\affil{Department of Physics and Astronomy, Aarhus University, Ny Munkegade, DK-8000 Aarhus C, Denmark}

\author[0000-0002-6650-694X]{K. Krisciunas}
\affil{George P. and Cynthia Woods Mitchell Institute for Fundamental Physics and Astronomy, Department of Physics and Astronomy, Texas A\&M University, College Station, TX, 77843, USA}

\author[0000-0001-8367-7591]{S. Kumar}
\affil{Department of Physics, Florida State University, 77 Chieftan Way, Tallahassee, FL 32306, USA}

\author[0000-0002-1132-1366]{H. Kuncarayakti}
\affil{Tuorla Observatory, Department of Physics and Astronomy, FI-20014 University of Turku, Finland}
\affil{Finnish Centre for Astronomy with ESO (FINCA), FI-20014 University of Turku, Finland}

\author[0000-0003-2535-3091]{N. Morrell}
\affil{Carnegie Observatories, Las Campanas Observatory, Colina El Pino, Casilla 601, Chile}

\author[0000-0003-1169-1954]{T. J. Moriya}
\affil{National Astronomical Observatory of Japan, National Institutes of Natural Sciences, 2-21-1 Osawa, Mitaka, Tokyo 181-8588, Japan}
\affil{School of Physics and Astronomy, Faculty of Science, Monash University, Clayton, Victoria 3800, Australia}

\author[0000-0002-3389-0586]{P. E. Nugent}
\affil{Lawrence Berkeley National Laboratory, Department of Physics, 1 Cyclotron Road, Berkeley, CA 94720, USA}
\affil{Astronomy Department, University of California at Berkeley, Berkeley, CA 94720, USA}

\author[0000-0002-4436-4661]{S. Perlmutter}
\affil{Lawrence Berkeley National Laboratory, Department of Physics, 1 Cyclotron Road, Berkeley, CA 94720, USA}
\affil{Astronomy Department, University of California at Berkeley, Berkeley, CA 94720, USA}

\author[0000-0003-0554-7083]{S. E. Persson}
\affil{Observatories of the Carnegie Institution for Science, 813 Santa Barbara St, Pasadena, CA 91101, USA}

\author[0000-0001-6806-0673]{A. L. Piro}
\affil{Observatories of the Carnegie Institution for Science, 813 Santa Barbara St, Pasadena, CA 91101, USA}

\author{D. Rabinowitz}
\affil{Physics Department, Yale University, 217 Prospect Street, New Haven, CT 06511, USA}

\author{M. Roth}
\affil{Carnegie Observatories, Las Campanas Observatory, Colina El Pino, Casilla 601, Chile}
\affil{GMTO Corporation, Avenida Presidente Riesco 5335, Suite 501, Las Condes, Santiago, Chile}

\author[0000-0002-9301-5302]{M. Shahbandeh}
\affil{Department of Physics, Florida State University, 77 Chieftan Way, Tallahassee, FL 32306, USA}

\author[0000-0003-4631-1149]{B. J. Shappee}
\affil{Institute for Astronomy, University of Hawaii, 2680 Woodlawn Drive, Honolulu, HI 96822,USA}

\author[0000-0002-5571-1833]{M. D. Stritzinger}
\affil{Department of Physics and Astronomy, Aarhus University, Ny Munkegade, DK-8000 Aarhus C, Denmark}

\author[0000-0002-8102-181X]{N. B. Suntzeff}
\affil{George P. and Cynthia Woods Mitchell Institute for Fundamental Physics and Astronomy, Department of Physics and Astronomy, Texas A\&M University, College Station, TX, 77843, USA}

\author[0000-0002-2387-6801]{F. Taddia}
\affil{Department of Physics and Astronomy, Aarhus University, Ny Munkegade, DK-8000 Aarhus C, Denmark}

\author[0000-0002-9413-4186]{S. A. Uddin}
\affil{Observatories of the Carnegie Institution for Science, 813 Santa Barbara St, Pasadena, CA 91101, USA}

\correspondingauthor{E. Y. Hsiao}
\email{ehsiao@fsu.edu}

\begin{abstract}
The Type Ia supernova (\snia) LSQ14fmg exhibits exaggerated properties
which may help to reveal the origin of the ``super-Chandrasekhar'' (or
03fg-like) group.  The optical spectrum is typical of a 03fg-like
\snia, but the light curves are unlike those of any \sneia\ observed.
The light curves of LSQ14fmg rise extremely slowly.  At $-23$
rest-frame days relative to $B$-band maximum, LSQ14fmg is already
brighter than $M_V=-19$ mag before host extinction correction.  The
observed color curves show a flat evolution from the earliest
observation to approximately one week after maximum.  The
near-infrared light curves peak brighter than $-20.5$ mag in the $J$
and $H$ bands, far more luminous than any 03fg-like \sneia\ with
near-infrared observations.  At one month past maximum, the optical
light curves decline rapidly.  The early, slow rise and flat color
evolution are interpreted to result from an additional excess flux
from a power source other than the radioactive decay of the
synthesized \nirad.  The excess flux matches the interaction with a
typical superwind of an asymptotic giant branch (AGB) star in density
structure, mass-loss rate, and duration.  \replaced{The late, rapid
  decline, soon after the supernova enters the presumed
  \corad\ radioactive decay tail}{The rapid decline starting at around
  one month past $B$-band maximum} may be an indication of rapid
cooling by active carbon monoxide (CO) formation, which requires a low
temperature and high density environment.  These peculiarities point
to an AGB progenitor near the end of its evolution and the core
degenerate scenario as the likely explosion mechanism for LSQ14fmg.
\end{abstract}


%%%%%%%%%%%%%%%%%%
%% Introduction %%
%%%%%%%%%%%%%%%%%%

\section{Introduction}
\label{s:intro}

%type Ia

Observations of distant Type Ia supernovae (\sneia) led to the
discovery of the accelerating expansion of the Universe
\citep{1998AJ....116.1009R, 1999ApJ...517..565P} and accurate
measurements of the Hubble-Lema\^{i}tre constant
\citep{2001ApJ...553...47F}.  Despite the success, there is no
consensus on the origin(s) of
\sneia\ \citep[e.g.,][]{2017ApJ...846...58H, 2017MNRAS.470..157B},
beyond that they are thermonuclear explosions of carbon-oxygen white
dwarfs \citep[C/O WDs;][]{1960ApJ...132..565H}.  The vast majority of
\sneia\ are remarkably uniform and follow the tight luminosity decline
rate relation \citep[or Phillips relation; ][]{1993ApJ...413L.105P}.
The bright-slow/faint-fast correlation enables \sneia\ to be used as
cosmological standard candles.  As observational data on
\sneia\ accumulate, objects with extreme or peculiar properties have
begun to emerge and form subgroups within the \snia\ classification,
such as the subluminous 91bg-like \citep{1992AJ....104.1543F,
  1993AJ....105..301L} and 02cx-like \citep{2003PASP..115..453L}, as
well as the overluminous 91T-like \citep{1992AJ....103.1632P,
  1992ApJ...384L..15F} and ``super-Chandrasekhar'' or 03fg-like
\citep{2006Natur.443..308H} subgroups \citep[for a review,
  see][]{2017hsn..book..317T}.  Due to their extreme nature, the
approach of identifying the origins of these subgroups which allows a
clearer definition of the normal population, has been shown to be more
fruitful than studying the normal objects alone
\citep[e.g.,][]{2003Natur.424..651H, 2013ApJ...767...57F,
  2014Natur.512...54M, 2017Natur.550...80J, 2019ApJ...873L..18D}.

%first super-C

The Chandrasekhar limit \citep{1931ApJ....74...81C} of
1.4~\msun\ provides the theoretical mass limit above which the
electron degeneracy pressure of a non-rotating WD can no longer
support the star against its gravity.  An exceptionally luminous
\snia\ discovered by the Supernova Legacy Survey \citep[SNLS-03D3bb or
  SN~2003fg;][]{2006Natur.443..308H} seemed to have defied this limit
and stands out from the normal population that follows the tight
Phillips relation.  Subsequent discoveries of the overluminous SNe
2007if \citep{2010ApJ...713.1073S, 2010ApJ...715.1338Y} and 2009dc
\citep{2009ApJ...707L.118Y, 2011MNRAS.410..585S, 2011MNRAS.412.2735T}
shared similar distinguishing properties and confirm this subgroup of
peculiar \sneia, commonly referred to as ``super-Chandrasekhar''
\sneia.

%naming challenge

As they are situated at nearly a full magnitude brighter than normal
\sneia\ with similar decline rates in the Phillips relation,
luminosity appeared to be the defining characteristic of this peculiar
subgroup.  However, the discoveries of fainter objects that share
other similarities with the overluminous ones, such as SNe~2006gz
\citep{2007ApJ...669L..17H}, 2012dn \citep{2014MNRAS.443.1663C,
  2016MNRAS.457.3702P, 2016PASJ...68...68Y, 2019MNRAS.488.5473T}, and
ASASSN-15pz \citep{2019ApJ...880...35C} suggests a wide range of
ejecta masses.  As the peak luminosities of the fainter objects extend
into the range of normal \sneia\ of similar decline rates, and other
power sources for their luminosities are suggested
\citep[e.g.,][]{2016MNRAS.463.2972N}, it is no longer clear that any
are ``super-Chandrasekhar'' in ejecta mass.  Thus, throughout this
paper, the term ``\superc'' will be used in place of
``super-Chandrasekhar,'' following the convention of naming the
peculiar subgroup after the first object of its kind discovered.  The
\superc\ objects are extremely rare.  To date, fewer than ten objects
have been identified as candidates, and a few more had marginal
identifying properties, such as SN~2004gu \citep{2010AJ....139..519C},
SN~2011hr \citep{2016ApJ...817..114Z}, LSQ12gdj
\citep{2014MNRAS.445...30S}, and iPTF13asv
\citep{2016ApJ...823..147C}, pointing to possible connections to other
peculiar subgroups.

%observational properties

The members of the \superc\ group have relatively low decline rates in
their $B$- and $V$-band light curves, positioning these \sneia\ on the
bright, slowly declining end of the Phillips relation.  They also have
slightly longer rise times compared to normal \sneia.  In general, the
optical spectra of \superc\ \sneia\ at maximum light resemble those of
normal objects, dominated by lines of intermediate-mass elements
(IMEs), such as Si, S, and Ca.  However, they also show distinguishing
characteristics: weak spectral lines and low expansion velocities at
maximum light (as low as $8{,}000$~\kms\ in the most luminous events).
Some \superc\ \sneia\ show evidence of substantial unburned material
via strong and/or persistent optical \cii\ features.  In the
near-infrared (NIR), on the other hand, the spectral features of
\superc\ and normal \sneia\ are drastically different
\citep{2019PASP..131a4002H}.  The prominent spectroscopic ``$H$-band
break'' which emerges a few days after maximum light in normal objects
\citep{1973ApJ...180L..97K, 1985ApJ...296..379E, 1998ApJ...496..908W}
is missing or appears much later in \superc\ objects
\citep{2011MNRAS.412.2735T, 2019PASP..131a4002H}.  Assuming that the
slower rise and high luminosity result from larger amounts of
\nirad\ produced, Arnett's rule \citep{1982ApJ...253..785A} yields
exceptionally high \nirad\ masses of $1.3-1.8$~\msun.  Even without
accounting for the IMEs and unburned material observed in the spectra,
some of these explosions are already near or above the Chandrasekhar
mass, unless there are other energy sources at play.

A normal \snia\ with high peak luminosity and slow decline in the $B$
band typically has an $i$-band light curve that peaks before the
$B$-band maximum and a prominent secondary maximum.  On the other
hand, all members of the \superc\ group have late $i$-band primary
maxima, occurring a few days after $B$-band maxima, and weak or no
secondary maxima in the $i$ band \citep{2014ApJ...795..142G,
  2020ApJ...895L...3A}.  The unique $i$-band morphology may be the
simplest way of distinguishing members of the \superc\ group from
normal or 91T-like objects.  So far, there have been only two
\superc\ objects with spectropolarimetric observations and both show
very low continuum polarization similar to normal \sneia: ${<}0.3$\%
in SN~2009dc \citep{2010ApJ...714.1209T} and $0.7$\% in SN~2007if
\citep{2019MNRAS.490..578C}.  For SN~2012dn, the NIR light curves were
observed to have sustained high luminosities, and the excess was
interpreted as an echo by circumstellar medium
\citep{2016PASJ...68...68Y}.  Assuming it is in an accreting WD
system, the continuum polarization is predicted to be as high as 8\%
\citep{2018MNRAS.476.4806N}.  There are also indications that
\superc\ events are systematically overluminous in the ultraviolet
compared to normal \sneia\ \citep{2014ApJ...787...29B}.  The
\superc\ objects tend to explode in low-mass, star-forming galaxies,
e.g., SNe~2003fg \citep{2006Natur.443..308H} and 2007if
\citep{2011ApJ...733....3C}.  If their host galaxies are more massive,
they tend to occur in more remote locations far from the centers of
their hosts, e.g., SNe~2009dc \citep{2011MNRAS.412.2735T} and 2012dn
\citep{2014MNRAS.443.1663C}, pointing to a preference for low
metallicity or young stellar population environments.

%theoretical explanations

Several theoretical explanations for the origin of the \superc\ group
have been proposed, but there is currently no single cohesive theory
that explains all the observations.  Differentially rotating WDs can
have masses well above the Chandrasekhar limit
\citep[e.g.,][]{1975ApJ...199..179D}.  However, the substantial
kinetic energy required to propagate the nuclear flame is in apparent
contradiction to the low photospheric velocities observed in the
brightest \superc\ events \citep{2012MNRAS.427.2057H}.  An off-center
explosion may result in nuclear burning toward a preferred direction
and an asymmetric distribution of \nirad\ synthesized
\citep[e.g.,][]{2007A&A...465L..17H}.  However, these models do not
reproduce the slow rise and decline in the observed light curves, but
produce substantial continuum polarization and shifts in the forbidden
Fe features in the nebular phase, in contradiction to observations.
The merger of two WDs has also been proposed as a possible explosion
scenario, either on secular \citep[e.g.,][]{1984ApJ...277..355W,
  2007MNRAS.380..933Y} or dynamical
\citep[e.g.,][]{2010Natur.463...61P, 2012ApJ...747L..10P} time scales,
although an accretion induced collapse \citep{1991ApJ...367L..19N} may
occur rather than a thermonuclear explosion.  The observed low
continuum polarization also appears to disfavor off-center explosions,
dynamical mergers, and even a differentially rotating WD
\citep{2003ApJ...595.1094U}.

The core degenerate scenario describes the merger of a WD and the
degenerate core of an asymptotic giant branch (AGB) star at or shortly
after the common envelope phase \citep[e.g.,][]{2003ApJ...594L..93L,
  2011MNRAS.417.1466K}.  The explosion may start as a deflagration or
a detonation \citep{2019nuco.conf..187H, 2019NewAR..8701535S}.  This
scenario should produce a spherical explosion that matches the low
continuum polarization observed, but spectral interaction signatures
between the ejecta and the hydrogen-rich AGB envelope have not been
observed.

%LSQ14fmg

In this paper, we present the Carnegie Supernova Project-II
\citep[CSP-II;][]{2019PASP..131a4001P} observations of the
\superc\ \snia\ LSQ14fmg.  This object exhibits several pronounced
peculiar properties which are unique or best observed among the
\superc\ group and may help to reveal the origin of \superc\ events.
The data set of LSQ14fmg and possible theoretical explanations for the
observed peculiarities are presented.  In Section~\ref{s:obs}, details
of the observations are described.  In Sections~\ref{s:phot} and
\ref{s:spec}, the photometric and spectroscopic properties of LSQ14fmg
are compared to those of normal and \superc\ \sneia.  In
Section~\ref{s:host}, an analysis of the host galaxy of LSQ14fmg is
presented.  Possible theoretical explanations for the unique observed
properties of LSQ14fmg are explored in Section~\ref{s:disc}, followed
by a summary of conclusions in Section~\ref{s:conc}.

%%%%%%%%%%%%%%%%%%
%% Observations %%
%%%%%%%%%%%%%%%%%%

\section{Observations}
\label{s:obs}

%discovery and classification

LSQ14fmg was discovered by the La Silla-QUEST Low Redshift Supernova
Survey \citep[LSQ;][]{2013PASP..125..683B} using an image taken on
2014 September 21.03 UT and confirmed with another image taken 2 hours
later.  Forced photometry on pre-discovery images revealed the
transient to be present as early as 2014 September 15.05 UT.  The
image of last non-detection was taken on 2014 August 20.21 UT, nearly
one month before the first detection.  The image yields a limit of
$21.9\pm0.1$ mag.  Stacked images from the All-Sky Automated Survey
for Supernovae \citep[ASAS-SN;][]{2014ApJ...788...48S,
  2017PASP..129j4502K} provided limits at four epochs during the time
gap between the last non-detection and the first detection.  The
Palomar Transient Factory \citep[PTF;][]{2009PASP..121.1395L} and the
Pan-STARRS imaging survey \citep{2010SPIE.7733E..0EK} did not cover
the field of LSQ14fmg during the time gap.  A classification spectrum
was taken on 2014 September 24.95 UT with the Andalucia Faint Object
Spectrograph and Camera (ALFOSC) on the Nordic Optical Telescope
(NOT).  LSQ14fmg was reported to be similar to a 91T-like \snia\ a few
days before maximum light due to its shallow \siii\ features
\citep{2014ATel.6495....1T}.

%photometric observations

\begin{figure}
\epsscale{1.1}
\plotone{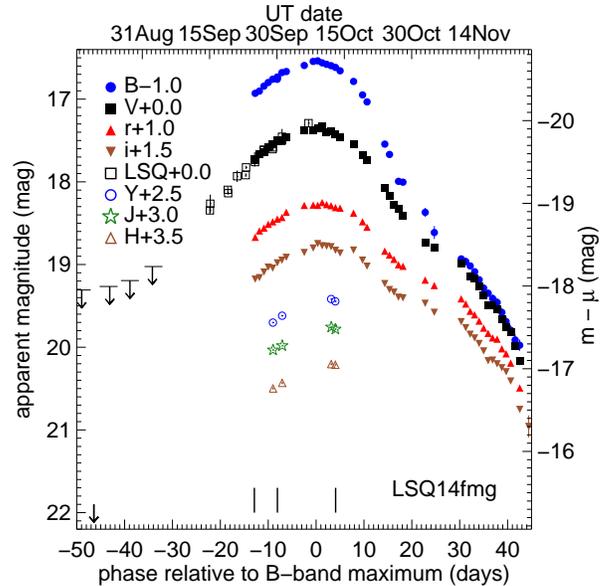}
\caption{Optical and NIR light curves of LSQ14fmg.  A non-detection
  limit from an image taken on 2014 August 20.21 UT by LSQ, as well as
  four less stringent limits from ASAS-SN, are represented by downward
  arrows.  The short vertical lines mark the dates when the follow-up
  spectra were taken.  The axis on the right is computed using a
  distance modulus of 37.26 mag.  The time axis is shown in the
  rest-frame.}
\label{f:lc}
\end{figure}

Since the supernova redshift, determined using the classification
spectrum, was well in the Hubble flow, LSQ14fmg was then followed up
by CSP-II as part of the ``Cosmology'' sample
\citep[see][]{2019PASP..131a4001P} in the optical $BVri$ bands with
the e2v CCD on the Swope Telescope and in the NIR $YJH$ bands with
RetroCam on the du~Pont Telescope.  The photometry was computed
relative to a local sequence of stars calibrated with respect to
standard star fields observed over typically 20 photometric nights in
the optical and three photometric nights in the NIR.  The 1-$\sigma$
uncertainties presented here correspond to the sum in quadrature of
the instrumental error and the nightly zero-point error.  The field of
LSQ14fmg was also monitored by the ESO 1-m Schmidt telescope and the
Quest camera of LSQ in the broad $gr$ filter with an approximately
$2$-day cadence until around maximum light.  The LSQ data were reduced
and the photometry was performed using the same methods as
\citet{2018ApJ...859...24C}.  The photometry is tabulated in the
natural photometric system of each of Swope+e2v, du~Pont+RetroCam, and
LSQ+QUEST in Tables~\ref{t:swope}, \ref{t:dupont}, and \ref{t:lsq},
respectively.  The color terms for transforming to the standard
systems are listed in \citet{2019PASP..131a4001P}.  Since the LSQ $gr$
band most closely resembles the $V$ band in terms of wavelength
coverage, an arbitrary zero point was added to the LSQ $gr$-band light
curve such that it matches the Swope $V$-band light curve.  No
S-corrections \citep{2002AJ....124.2100S} were included.  All the
light curves have had the host galaxy light removed using host galaxy
templates taken before the explosion (LSQ) or $\sim200$ days ($BVri$)
to $\sim300$ ($YJH$) after maximum light.  These light curves are
presented in Fig.~\ref{f:lc}.

%spectroscopic observations

\begin{figure}
\epsscale{1.1}
\plotone{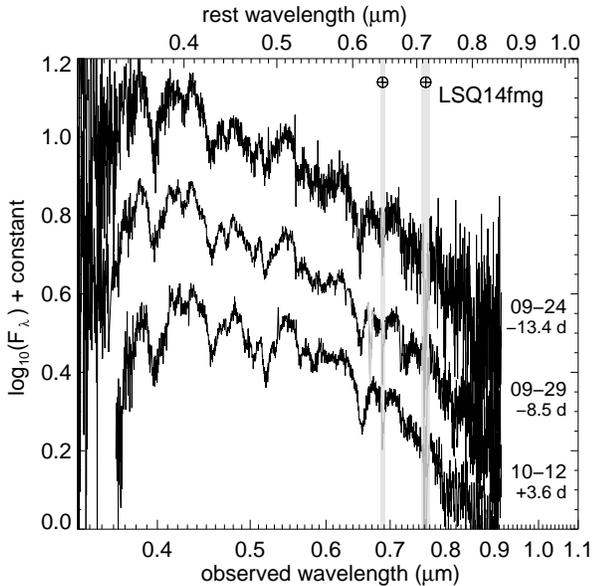}
\caption{Optical spectra of LSQ14fmg.  The UT date of the observation
  and rest-frame phase relative to $B$-band maximum are labeled for
  each spectrum.  The gray vertical bands mark the regions of telluric
  absorptions.  A sharp feature in the second spectrum near observer
  frame 0.67 \um\ is an artifact from the second-order contamination
  correction and is thus grayed out.  There is very little spectral
  evolution during the $\sim2.5$-week coverage.}
\label{f:spec}
\end{figure}

After the first classification spectrum was taken with NOT+ALFOSC, two
additional follow-up spectra were also obtained with the NOT, creating
a time series that spans approximately 2.5 weeks.  All spectra were
reduced in the standard manner using \texttt{IRAF}\footnote{The Image
  Reduction and Analysis Facility (IRAF) is distributed by the
  National Optical Astronomy Observatory, which is operated by the
  Association of Universities for Research in Astronomy, Inc., under
  cooperative agreement with the National Science Foundation.}
scripts.  A journal of the spectroscopic observations is presented in
Table~\ref{t:not}, and the three spectra are shown in
Fig.~\ref{f:spec}.

%host galaxy

\begin{figure}
\epsscale{1.1}
\plotone{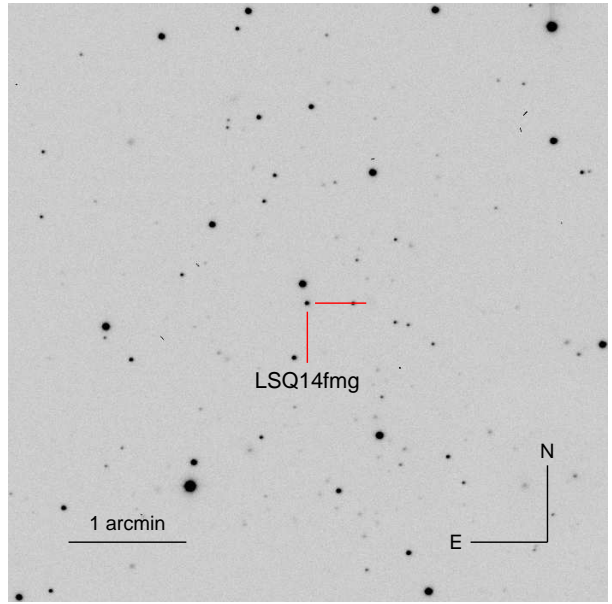}
\caption{The first Swope $i$-band image of LSQ14fmg taken
  approximately two weeks before $B$ maximum.  The crosshair marks the
  location of the supernova.  The compass and the size of the field
  are also noted.}
\label{f:image}
\end{figure}

The first $i$-band image taken with the Swope Telescope is shown in
Fig.~\ref{f:image}.  The host galaxy of LSQ14fmg is not immediately
apparent in the follow-up images.  However, the site of LSQ14fmg is in
the Sloan Digital Sky Survey (SDSS) Data Release 15 footprint
\citep{2019ApJS..240...23A}.  The center of a low-luminosity galaxy,
SDSS~J221646.15+152114.2 coincides with the position of LSQ14fmg
within 1\arcsec.  We therefore assume this is the host galaxy of
LSQ14fmg.  After the supernova had faded, on 2015 July 19.03 UT, a
spectrum of the host galaxy was taken with Wide-Field CCD (WFCCD)
spectrograph on the du\,Pont telescope.  A host galaxy heliocentric
redshift of $z_{\mathrm{helio}}=0.0661\pm0.0003$ was measured using
four narrow emission lines of H$\alpha$, [\ion{N}{2}] and
[\ion{S}{2}].  Correcting to the reference frame defined by the cosmic
microwave background (CMB) radiation \citep{1996ApJ...473..576F}, the
redshift becomes $z_{\mathrm{CMB}}=0.0649$.  This corresponds to a
distance modulus of $37.26\pm0.03$, assuming
$H_0=72$~\kms\,Mpc$^{-1}$.  The distance to the host galaxy of
LSQ14fmg is known quite precisely due to the small uncertainty
contribution from any peculiar velocities.  The properties of the host
galaxy are summarized in Table~\ref{t:prop}.

%host IFS observations

Integral field spectroscopy of the LSQ14fmg host galaxy was obtained
on 2017 August 4, around three years after the supernova explosion,
with the Multi-Unit Spectroscopic Explorer
\citep[MUSE;][]{2010SPIE.7735E..08B}, mounted to the Unit 4 telescope
(UT4) at the Very Large Telescope (VLT) of the Cerro Paranal
Observatory.  The observations were obtained as part of the
All-weather MUse Supernova Integral-field Nearby Galaxies
\citep[AMUSING;][]{2016MNRAS.455.4087G} survey, aimed at studying the
host environments of a large sample of nearby supernovae.  The
wide-field mode of MUSE provides a field-of-view of approximately
1\arcmin$\times$1\arcmin\ and a squared spatial pixel of
0.2\arcsec\ per side, which limits the spatial resolution.  The
atmospheric seeing during the observations was measured to be
1.16\arcsec.  The wavelength coverage ranges from 4750 to 9300 \AA,
with a spectral resolution from $\lambda$/$\Delta\lambda$ $\sim$ 1800
on the blue end to 3600 on the red end of the spectrum.  A detailed
explanation of the data reduction is provided in
\cite{2017A&A...602A..85K}.  Briefly, version 1.2.1 of the MUSE
reduction pipeline \citep{2014ASPC..485..451W} and the \texttt{Reflex}
environment \citep{2013A&A...559A..96F} were used.  The sky
subtraction was performed using the Zurich Atmosphere Purge package
(ZAP; \citealt{2016MNRAS.458.3210S}), employing blank sky regions
within the science frame.  The effects of Galactic extinction were
also corrected based on the reddening estimates from
\cite{2011ApJ...737..103S}.  Analysis of this data is presented in
Sec.~\ref{s:host}.

\begin{deluxetable}{ccccc}
\tablecolumns{5}
\tablecaption{Swope+e2v photometry of LSQ14fmg in the natural system.\label{t:swope}}
\tabletypesize{\scriptsize}
\tablehead{
\colhead{MJD} &
\colhead{$B$} &
\colhead{$V$} &
\colhead{$r$} &
\colhead{$i$}
}
\startdata
56925.15 & 17.930 (0.012) & 17.736 (0.012) & 17.671 (0.010) & 17.677 (0.013) \\
56926.08 & 17.903 (0.008) & 17.671 (0.009) & 17.593 (0.008) & 17.659 (0.011) \\
56927.10 & 17.843 (0.011) & 17.638 (0.010) & 17.559 (0.009) & 17.591 (0.012) \\
56928.05 & 17.799 (0.010) & 17.586 (0.008) & 17.513 (0.009) & 17.532 (0.010) \\
56929.05 & 17.759 (0.009) & 17.544 (0.010) & 17.483 (0.008) & 17.544 (0.010) \\
56930.07 & 17.760 (0.032) & 17.505 (0.008) & 17.450 (0.008) & 17.480 (0.010) \\
56931.10 & 17.679 (0.012) & 17.491 (0.018) & 17.431 (0.013) & 17.448 (0.015) \\
56932.07 & 17.666 (0.010) & 17.456 (0.009) & 17.367 (0.010) & 17.413 (0.012) \\
56936.08 & 17.593 (0.018) & 17.381 (0.014) & 17.285 (0.013) & 17.355 (0.014) \\
56938.06 & 17.544 (0.025) & 17.378 (0.019) & 17.281 (0.019) & 17.301 (0.020) \\
56939.07 & 17.537 (0.019) & 17.352 (0.020) & 17.284 (0.018) & 17.247 (0.015) \\
56940.06 & 17.561 (0.013) & 17.323 (0.016) & 17.250 (0.013) & 17.273 (0.013) \\
56941.05 & 17.579 (0.009) & 17.396 (0.008) & 17.275 (0.008) & 17.278 (0.011) \\
56942.04 & 17.595 (0.008) & 17.384 (0.010) & 17.288 (0.009) & 17.285 (0.011) \\
56943.03 & 17.616 (0.011) & 17.426 (0.009) & 17.311 (0.010) & 17.328 (0.014) \\
56944.05 & 17.657 (0.017) & 17.455 (0.017) & 17.318 (0.011) & 17.361 (0.015) \\
56947.05 & 17.786 (0.010) & 17.549 (0.011) & 17.378 (0.014) & 17.326 (0.015) \\
56949.05 & 17.949 (0.010) & 17.680 (0.011) & 17.483 (0.009) & 17.447 (0.010) \\
56950.05 & 18.033 (0.015) & 17.736 (0.012) & 17.549 (0.011) & 17.520 (0.011) \\
56954.01 & 18.544 (0.017) & 18.080 (0.012) & 17.838 (0.010) & 17.731 (0.013) \\
56955.05 & 18.669 (0.022) & 18.170 (0.014) & 17.885 (0.012) & 17.803 (0.012) \\
56956.05 &       $\cdots$ & 18.280 (0.033) & 17.937 (0.013) & 17.834 (0.015) \\
56957.06 & 18.994 (0.018) & 18.329 (0.014) & 18.003 (0.012) & 17.895 (0.014) \\
56958.05 & 19.004 (0.019) & 18.411 (0.015) & 18.021 (0.011) & 17.903 (0.013) \\
56963.00 & 19.369 (0.054) & 18.739 (0.038) & 18.188 (0.025) & 17.969 (0.022) \\
56965.03 & 19.613 (0.078) & 18.800 (0.041) & 18.254 (0.020) & 18.078 (0.018) \\
56971.03 & 19.932 (0.039) & 18.986 (0.031) & 18.415 (0.016) & 18.191 (0.016) \\
56972.04 & 19.963 (0.044) &       $\cdots$ & 18.477 (0.014) & 18.264 (0.017) \\
56973.03 & 20.017 (0.027) & 19.147 (0.021) & 18.565 (0.016) & 18.343 (0.017) \\
56974.02 & 20.087 (0.029) & 19.174 (0.020) & 18.608 (0.018) & 18.388 (0.017) \\
56975.02 & 20.184 (0.029) & 19.259 (0.022) & 18.684 (0.019) & 18.455 (0.025) \\
56976.02 & 20.285 (0.042) & 19.372 (0.024) & 18.769 (0.020) & 18.545 (0.021) \\
56977.02 & 20.347 (0.042) & 19.491 (0.031) & 18.832 (0.022) & 18.663 (0.029) \\
56978.02 & 20.411 (0.050) & 19.498 (0.028) & 18.887 (0.020) & 18.656 (0.017) \\
56979.02 & 20.458 (0.053) & 19.545 (0.031) & 18.902 (0.019) & 18.701 (0.024) \\
56980.08 & 20.579 (0.055) & 19.667 (0.034) & 19.019 (0.020) & 18.748 (0.026) \\
56981.03 & 20.690 (0.057) & 19.762 (0.030) & 19.079 (0.018) & 18.797 (0.025) \\
56982.05 & 20.795 (0.058) & 19.819 (0.031) & 19.192 (0.021) & 18.909 (0.023) \\
56983.04 & 20.912 (0.070) & 19.991 (0.036) &       $\cdots$ &       $\cdots$ \\
56984.06 & 20.976 (0.068) & 20.168 (0.048) & 19.493 (0.024) & 19.252 (0.030) \\
56986.06 &       $\cdots$ &       $\cdots$ &       $\cdots$ & 19.458 (0.134) \\
56994.05 &       $\cdots$ &       $\cdots$ &       $\cdots$ & 19.705 (0.100) \\
56996.05 &       $\cdots$ &       $\cdots$ &       $\cdots$ & 20.037 (0.132) \\
\enddata
%\tablecomments{}
\end{deluxetable}

\begin{deluxetable}{cccc}
\tablecolumns{4}
\tablecaption{du\,Pont+RetroCam photometry of LSQ14fmg in the natural system.\label{t:dupont}}
\tabletypesize{\scriptsize}
\tablehead{
\colhead{MJD} &
\colhead{$Y$} &
\colhead{$J$} &
\colhead{$H$}
}
\startdata
56929.11 & 17.202 (0.009) & 17.033 (0.012) & 16.999 (0.014) \\
56931.16 & 17.119 (0.011) & 16.981 (0.010) & 16.932 (0.016) \\
56942.08 & 16.917 (0.010) & 16.756 (0.011) & 16.705 (0.015) \\
56942.99 & 16.945 (0.011) & 16.781 (0.012) & 16.712 (0.017) \\
\enddata
%\tablecomments{}
\end{deluxetable}

\begin{deluxetable}{cc}
\tablecolumns{2}
\tablecaption{LSQ+QUEST photometry of LSQ14fmg in the natural system.\label{t:lsq}}
\tabletypesize{\scriptsize}
\tablehead{
\colhead{MJD} &
\colhead{LSQ $gr$}
}
\startdata
56915.05 & 18.3453 (0.042) \\
56915.13 & 18.2650 (0.110) \\
56919.04 & 18.0967 (0.039) \\
56919.12 & 18.1374 (0.018) \\
56921.12 & 17.9322 (0.069) \\
56923.03 & 17.9187 (0.022) \\
56923.11 & 17.8252 (0.019) \\
56925.02 & 17.7274 (0.088) \\
56925.10 & 17.7614 (0.029) \\
56927.02 & 17.6167 (0.030) \\
56927.10 & 17.6105 (0.036) \\
56929.02 & 17.6052 (0.018) \\
56929.11 & 17.5840 (0.035) \\
56931.01 & 17.5045 (0.013) \\
56931.12 & 17.4551 (0.095) \\
56937.02 & 17.2903 (0.045) \\
\enddata
%\tablecomments{}
\end{deluxetable}

\begin{deluxetable}{cccc}
\tablecolumns{4}
\tablecaption{Journal of spectroscopic observations.\label{t:not}}
\tabletypesize{\scriptsize}
\tablehead{
\colhead{UT date} &
\colhead{MJD} &
\colhead{Instrument} &
\colhead{$t_{\rm{max}}(B)$\tablenotemark{a}}
}
\startdata
2014-09-24 & 56924.95 & NOT+ALFOSC & $-13.4$ \\ %updated 2020-06-23
2014-09-29 & 56930.10 & NOT+ALFOSC & $ -8.5$ \\
2014-10-12 & 56943.06 & NOT+ALFOSC & $ +3.6$ \\ 
\enddata
\tablenotetext{a}{Rest-frame days relative to $B$-band maximum.}
\end{deluxetable}

\begin{deluxetable}{lc}
\tablecolumns{2}
\tablecaption{Properties of LSQ14fmg and its presumed host galaxy.\label{t:prop}}
\tabletypesize{\footnotesize}
\tablehead{
}
\startdata
LSQ14fmg \\
$\alpha (J2000)$ & 334.192125\arcdeg \\
$\delta (J2000)$ & +15.353925\arcdeg \\
\dm    & $1.062\pm0.058$ mag \\
\sbv   & $1.180\pm0.067$ \\
JD$_{\mathrm{max}}(B)$ & $2456939.7\pm0.5$ \\ %%Jing k-/MW corrected, 2020-06-23
JD$_{\mathrm{max}}(V)$ & $2456938.5\pm0.5$ \\
JD$_{\mathrm{max}}(r)$ & $2456940.7\pm0.7$ \\
JD$_{\mathrm{max}}(i)$ & $2456940.7\pm0.4$ \\
$B_{\mathrm{max}}$\tablenotemark{a} & $17.385\pm0.008$ mag \\
$V_{\mathrm{max}}$\tablenotemark{a} & $17.273\pm0.008$ mag \\
$r_{\mathrm{max}}$\tablenotemark{a} & $17.175\pm0.005$ mag \\
$i_{\mathrm{max}}$\tablenotemark{a} & $17.256\pm0.006$ mag \\
$Y_{\mathrm{max}}$\tablenotemark{b} & $16.862\pm0.010$ mag \\
$J_{\mathrm{max}}$\tablenotemark{b} & $16.715\pm0.011$ mag \\
$H_{\mathrm{max}}$\tablenotemark{b} & $16.678\pm0.015$ mag \\
$M_{B\mathrm{,max}}$\tablenotemark{c} & $-19.87\pm0.03$ mag \\
$M_{V\mathrm{,max}}$\tablenotemark{c} & $-19.99\pm0.03$ mag \\
$M_{r\mathrm{,max}}$\tablenotemark{c} & $-20.08\pm0.03$ mag \\
$M_{i\mathrm{,max}}$\tablenotemark{c} & $-20.00\pm0.03$ mag \\
$M_{Y\mathrm{,max}}$\tablenotemark{d} & $-20.40\pm0.03$ mag \\
$M_{J\mathrm{,max}}$\tablenotemark{d} & $-20.55\pm0.03$ mag \\
$M_{H\mathrm{,max}}$\tablenotemark{d} & $-20.58\pm0.03$ mag \\
$M_{B\mathrm{,max}}$\tablenotemark{e} & $-20.3\pm0.5$ mag \\
\hline
SDSS J221646.15+152114.2 \\
$\alpha (J2000)$ & 334.192306\arcdeg \\
$\delta (J2000)$ & +15.353950\arcdeg \\
$z_{\mathrm{helio}}$ & $0.0661\pm0.0003$ \\
$z_{\mathrm{CMB}}$ & $0.0649$ \\
$r$ & $22.40\pm0.53$ mag \\
$M_r$ & $-14.86\pm0.53$ mag \\
distance modulus & $37.26\pm0.03$ mag \\
E($B-V$)$_{\mathrm{host}}$ & $0.13\pm0.15$ mag \\
stellar mass & 1.15$\pm$0.26$\times$10$^9$ M$_\odot$ \\
SFR & 0.081$\pm$0.027 M$_\odot$ yr$^{-1}$ \\
$\log_{10}$(sSFR[yr$^{-1}$]) & $-10.14\pm$0.49 \\
\enddata
\tablenotetext{a}{Apparent magnitude without any corrections.}
\tablenotetext{b}{Apparent magnitude of the brightest light-curve point including Milky Way extinction correction.}
\tablenotetext{c}{Absolute magnitude including $K$-correction and Milky Way extinction correction.}
\tablenotetext{d}{Absolute magnitude of the brightest light-curve point including Milky Way extinction corrections.}
\tablenotetext{e}{Absolute magnitude including $K$-correction, Milky Way, and host extinction corrections.}
%\tablecomments{}
\end{deluxetable}

%%%%%%%%%%%%%%%%
%% Photometry %%
%%%%%%%%%%%%%%%%

\section{Photometric properties}
\label{s:phot}

%intro

Although there is no doubt that LSQ14fmg is a \snia\ from its spectra,
the light curves do not resemble those of a normal or
\superc\ \snia. In this section, the peculiar photometric properties
of LSQ14fmg are explored.

%host extinction

For a peculiar and unique transient like LSQ14fmg, it is difficult to
derive its host extinction from the light curves, since the intrinsic
colors are unknown.  A \naid\ absorption feature at the host redshift
is detected in the last two low-resolution spectra, which have higher
signal-to-noise ratios than the classification spectrum.  The
equivalent widths (EWs) of the \naid\ absorption of the two spectra
spanning approximately two weeks are consistent with each other.
Taking the value of the last spectrum, we obtained $0.82\pm0.38$ \AA.
Adopting the empirical relation from \citet{2012MNRAS.426.1465P},
while including the measurement error and the dispersion around the
empirical relation \citep{2013ApJ...779...38P}, yields a very
uncertain color excess of E$(B-V)_{\mathrm{host}}=0.13\pm0.16$ mag.
The extinction estimate from the Balmer decrement of the MUSE
observation at the location of the supernova is also consistent with
zero.

%peak absolute magnitude

\begin{figure}
\epsscale{0.9}
\plotone{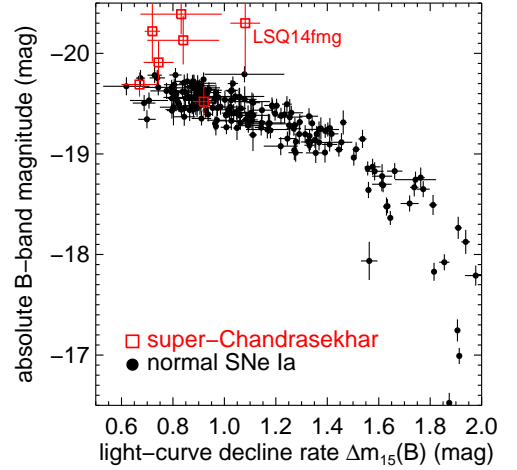}
\caption{The Phillips relation constructed using \sneia\ observed by
  the CSP \citep{2017AJ....154..211K, 2019PASP..131a4001P}.  Other
  published \superc\ objects and LSQ14fmg in particular are
  highlighted for comparison.}
\label{f:wlr}
\end{figure}

Peak brightness is a key diagnostic of \superc\ objects.  The $B$-band
peak apparent magnitude of LSQ14fmg without any correction is
$17.385\pm0.008$ mag.  Applying $K$-corrections, Milky Way extinction
correction, and the distance modulus yields an absolute magnitude of
$-19.87\pm0.03$ mag in the $B$ band.  Without correcting for host
extinction, the peak absolute $B$ magnitude of LSQ14fmg lies in the
middle of the range for \superc\ objects \citep{2017hsn..book..317T}.
Applying the uncertain host extinction correction while assuming a
total-to-selective extinction ratio of $R_V=3.1$, then brings the peak
absolute $B$ magnitude to $-20.3\pm0.5$ mag, one of the brightest in
the \superc\ group (Fig.~\ref{f:wlr}).  The photometric properties of
LSQ14fmg are summarized in Table~\ref{t:prop}.  \added{These
  quantities were derived using light curves of LSQ14fmg that are in
  the CSP natural system.}

%k-correction

Since there are only three observed spectra of LSQ14fmg, the
time-series spectra of SN~2009dc \citep{2011MNRAS.410..585S,
  2011MNRAS.412.2735T} were used to calculate the $K$-corrections.
The spectra of SN~2009dc were warped \citep{2007ApJ...663.1187H} to
match the observed colors of LSQ14fmg before the $K$-corrections were
computed.  The uncertainty associated with this approach is measured
to be less than 0.03 mag.  Throughout the paper, the $B$-band maximum
JD date of $2456939.7\pm0.5$, measured using the $K$-corrected and
Milky Way extinction corrected light curve, was adopted.  Note that
the $K$-corrected $B$-band light curve peaks slightly later than the
uncorrected one published in \citet{2020ApJ...895L...3A}, who analyzed
a large sample of low-redshift peculiar objects.  For other
\superc\ objects used in this paper, time-series spectra of SNe~2009dc
\citep{2011MNRAS.410..585S, 2011MNRAS.412.2735T}, 2012dn
\citep{2019MNRAS.488.5473T}, or spectral templates of
\citet{2007ApJ...663.1187H} were used, depending on which data set
provided the closest match to the $K$-corrections computed with
observed spectra.  These time-series spectroscopic data sets were
chosen for their short cadence and extensive coverage in phase.  For
the remainder of the analysis, we use all $K$-corrected and Milky Way
extinction corrected light curves, except for the LSQ+QUEST $gr$ light
curve of LSQ14fmg, which is artificially shifted to match the $V$-band
light curve.

%natural system

\added{As much as possible, comparison light curves were placed in the
  CSP natural system, since the color terms derived using standard
  stars may not be appropriate for supernova magnitudes
  \citep{2017AJ....154..211K}.  For data published by the CfA
  Supernova Group in their natural system, the light curves were
  simply S-corrected to the CSP system.  For observations taken with
  SNIFS \citep{2004SPIE.5249..146L}, the spectra were convolved with
  CSP filter functions to obtain the light curves.  For the remaining
  data sets, they are left in the standard system as published.}

%LC comparison data

\begin{figure*}
\centering
\epsscale{1.1}
\plottwo{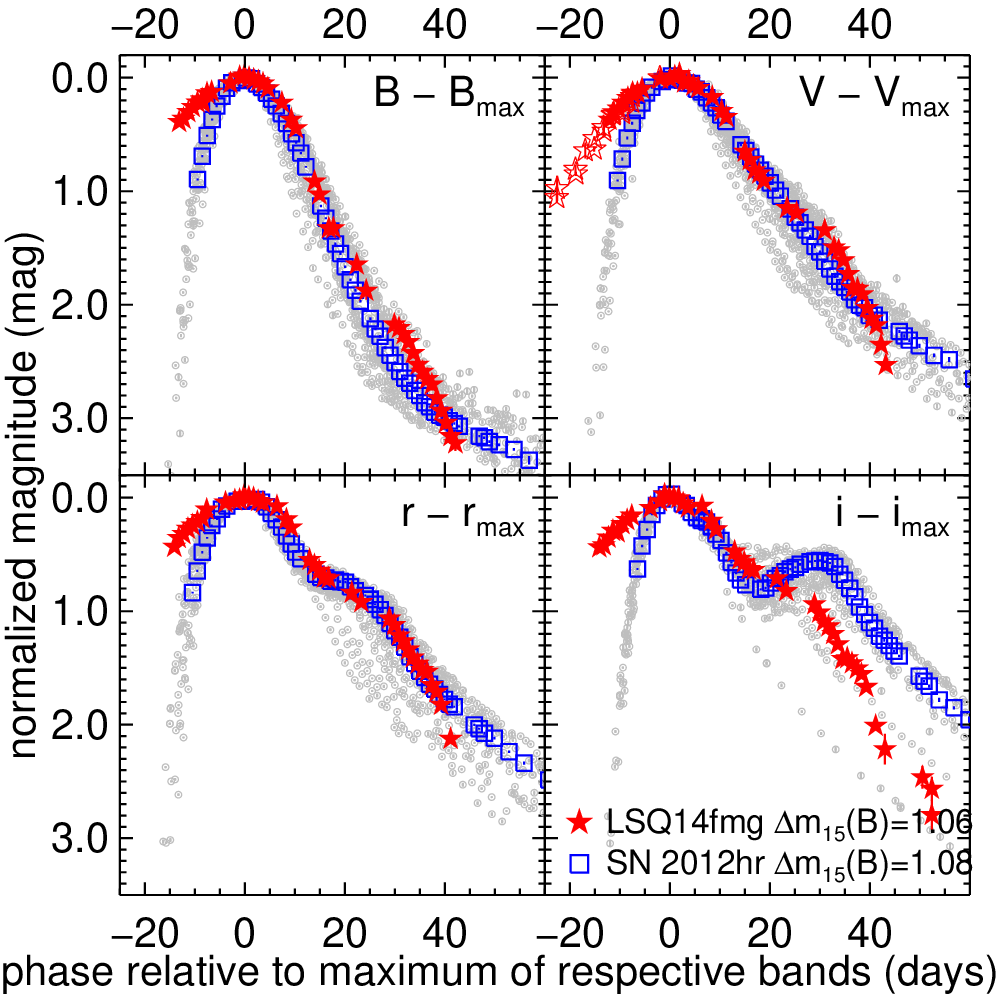}{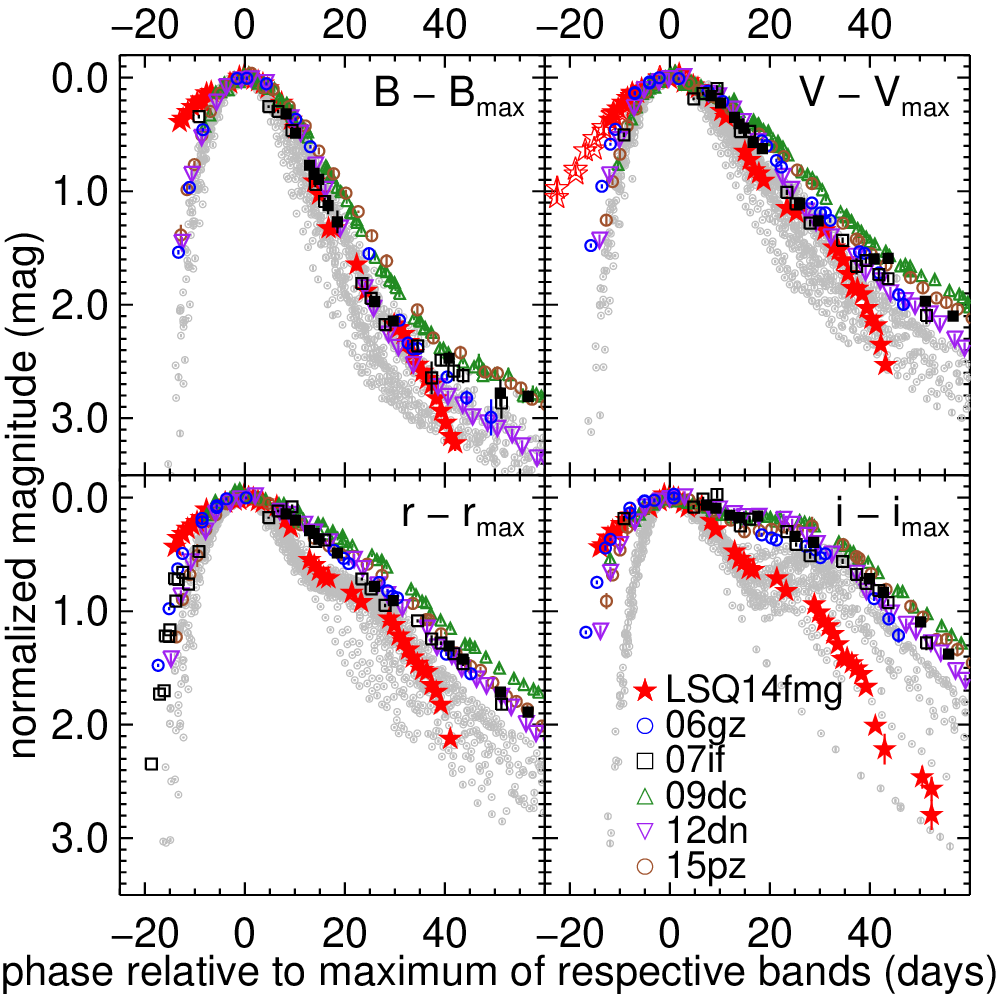}
\caption{Light-curve shape comparisons between LSQ14fmg and a normal
  \snia\ with a similar decline rate (left panels) and all published
  low-redshift \superc\ \sneia\ (right panels).  All light curves are
  shifted in both the time and magnitude axes to match the peak
  magnitude in each band.  The time axis is shown in the rest-frame of
  each \snia.  \sneia\ from CSP-I and CSP-II that are well-sampled and
  cover extended phase range are plotted as gray circles in the
  background.  The light curves of LSQ14fmg are plotted as filled red
  stars.  The LSQ $gr$ light curve is plotted with the $V$ band as
  open red stars.}
\label{f:LC_comp}
\end{figure*}

In Fig.~\ref{f:LC_comp}, the light-curve shape of LSQ14fmg is compared
with a normal \snia\ and other \superc\ objects.  In the left panel,
normal \snia\ SN~2012hr from the CSP-II sample
\citep{2019PASP..131a4001P} was chosen since it has a very similar
\dm\ value compared to LSQ14fmg.  In the right panel, the light curves
of LSQ14fmg are compared with those of SNe~2006gz
\citep{2007ApJ...669L..17H}, 2007if \citep{2010ApJ...713.1073S,
  2010ApJ...715.1338Y, 2017AJ....154..211K}, 2009dc
\replaced{\citep{2011MNRAS.412.2735T,
    2017AJ....154..211K}}{\citep{2011MNRAS.412.2735T,
    2012ApJS..200...12H, 2017AJ....154..211K}}, 2012dn
\citep{2019MNRAS.488.5473T}, and ASASSN-15pz
\citep{2019ApJ...880...35C}.  All light curves are $K$-corrected
except for the early LSQ $gr$ light curve of LSQ14fmg and the early
ROTSE-III unfiltered light curve of SN~2007if adjusted to SDSS $r$ as
described by \citet{2010ApJ...715.1338Y}.  All light curves are also
corrected for Milky Way extinction and time dilation.  Filled symbols
represent CSP data.

%rise

The rise of LSQ14fmg in all bands is strikingly slow compared to
normal \sneia.  The light-curve shapes of LSQ14fmg are compared with
those of normal \sneia\ from CSP-I \citep{2017AJ....154..211K} and
CSP-II \citep{2019PASP..131a4001P} in Fig.~\ref{f:LC_comp}.  Members
of the \superc\ group have slightly longer rise times to maximum,
e.g., 24 days in SN~2007if \citep{2010ApJ...713.1073S}, compared to
the $\sim$19-day rise time of the normal population
\citep[e.g.,][]{2006AJ....132.1707C}.  The rise time of LSQ14fmg may
be much longer than any \sneia, and it certainly has the slowest rise
ever observed in a \snia\ during the time segment from the first data
point to maximum.  The first light-curve point captured by LSQ is at
22.6 rest-frame days before $B$ maximum and is already as bright as
$M_V=-19.1\pm0.1$ mag before host extinction correction
(Fig.~\ref{f:lc}).  Unfortunately, the last LSQ non-detection was more
than 25 days before the first LSQ point, preventing an accurate
estimate of the rise time (Section~\ref{s:obs}).  For an extremely
bright object like SN~2009dc, the rise in the $B$ and $V$ bands is
slower than most normal \sneia, perhaps due to a larger amount of
\nirad\ produced, but is clearly still much faster than that of
LSQ14fmg (right panels of Fig.~\ref{f:LC_comp}).  Early ROTSE-III data
\citep{2010ApJ...715.1338Y} in combination with SNIFS data
\citep{2010ApJ...713.1073S} in the $r$ band showed that SN~2007if,
another extremely luminous \superc\ object, also has a much faster
rise than LSQ14fmg (right panels of Fig.~\ref{f:LC_comp}).  The above
suggests that the extremely slow rise of LSQ14fmg is unique and the
high pre-maximum luminosity may have a power source other than \nirad.

%BVr bands

The post-maximum decline is not unusual in the $B$, $V$, and $r$
bands, and the decline rate lies in the slower end of this particular
sample of normal \sneia\ (Fig.~\ref{f:LC_comp}).  The decline rate of
LSQ14fmg as defined by \citet{1993ApJ...413L.105P} is directly
measured to be \dm\,$=1.062\pm0.058$ mag and the color stretch
parameter \citep{2014ApJ...789...32B} \sbv\,$=1.180\pm0.067$, which
are not extreme.  In the left panels of Fig.~\ref{f:LC_comp}, the
light curves of a normal \snia, SN~2012hr, with a similar decline
rate, \dm=$1.075\pm0.009$ mag, are highlighted for comparison.  The
post-maximum evolution is nearly identical between SN~2012hr and
LSQ14fmg in the $BVr$ bands until \replaced{the radioactive decay
  tail}{around one month past $B$-band maximum}.  The likeness to
normal \sneia\ in the bluer bands, in combination with the preference
for low-luminosity and star-forming hosts (Section~\ref{s:host}),
means that an object like LSQ14fmg can be a significant contaminant to
a high-redshift cosmology \snia\ sample, which may only sample the
bluer rest-frame bands at a few epochs.  All \superc\ \sneia\ have
normal $B$-band post-maximum decline, with decline rates at the slower
end of the normal \snia\ population.  The differences in the
post-maximum evolution between normal and \superc\ objects become
larger from blue to red filter bands (right panel of
Fig.~\ref{f:LC_comp}) as pointed out by previous studies
\citep[e.g.,][]{2014ApJ...795..142G}.

%i bands

The largest photometric differences in the optical can be seen in the
$i$ band (Fig.~\ref{f:LC_comp}).  A \snia\ with high peak luminosity
and relatively slow decline in the $B$ band is typically accompanied
by a prominent secondary maximum in the $i$ band
\citep[e.g.,][]{2010AJ....139..120F}.  In addition, the primary
maximum of the $i$-band typically peaks before that of $B$ band.  On
the contrary, LSQ14fmg has a weak $i$-band secondary maximum and has
an $i$-band primary maximum that peaks slightly after $B$-band
maximum, properties which resemble that of a fast-declining,
subluminous \snia.  All members of the \superc\ group have much weaker
or no secondary maxima in the $i$ band and have the primary maxima of
the $i$ band peaking after those in the $B$ band
\citep{2020ApJ...895L...3A}.  Indeed, the timing of the $i$-band
primary maxima and the strength of the $i$-band secondary maxima of
objects with slow-declining $B$-band light curves may be the simplest
way of distinguishing members of the \superc\ group from
slow-declining normal \sneia\ or 91T-like objects.  SN~2004gu was
noted to be similar in peak brightness, decline rate, and spectral
features to SN~2006gz by \citet{2010AJ....139..519C}.  However, the
$iYJH$-band light curves of SN~2004gu all show strong secondary
maxima.  While SN~2004gu may be physically related to the
\superc\ subgroup, observationally it shares more similarities with
the 91T-like subgroup.  It is thus not treated as an \superc\ event
here.

%decay tail drop

Another peculiarity of LSQ14fmg surfaces \replaced{as the $B$ and $V$
  light curves reach the presumed radioactive decay tail}{at around
  one month past $B$-band maximum}.  The $B$ and $V$ light curves of
normal, as well as \superc\ \sneia, flatten, while those of LSQ14fmg
flatten briefly around the same epoch, but then the decline rate
increases rapidly around one month past maximum
(Fig.~\ref{f:LC_comp}).  The unusually rapid decline is also seen to
some degree in the $r$ and $i$ bands.  The rapid dimming \replaced{at
  the start of the presumed radioactive decay tail}{starting at around
  one month past maximum} is a unique characteristic of LSQ14fmg even
compared to the peculiar \superc\ group.  Note, however, that a few
\superc\ \sneia\ show evidence of a period of more rapid decline than
normal \sneia, but at a much later epoch, beyond 70 days past maximum
in SN~2012dn \citep{2014MNRAS.443.1663C} and beyond 200 days past
maximum in SN~2009dc \citep{2013MNRAS.432.3117T, 2009ApJ...690.1745M}.
Other \superc\ may not have adequate late-time light curve coverage to
characterize this dimming.  This peculiar feature may be shared among
\superc\ objects and caused by a common mechanism.

%colors

\begin{figure}
\epsscale{1.1}
\plotone{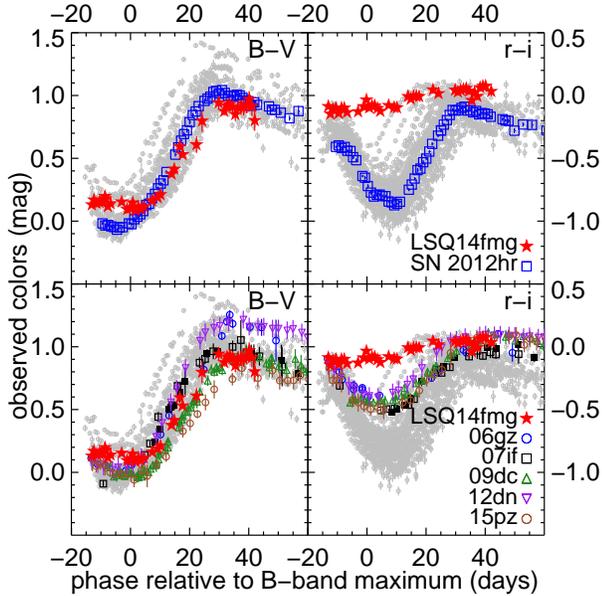}
\caption{The observed $B-V$ (left panels) and $r-i$ (right panels)
  colors of LSQ14fmg.  The colors are plotted in comparison to normal
  \sneia\ from CSP-I and CSP-II as gray backgrounds, and to SN~2012hr,
  a normal \snia\ of a similar \dm\ value (top panels), and to other
  \superc\ objects (bottom panels).  These plots use the same data set
  shown in Fig.~\ref{f:LC_comp} and are without host extinction
  corrections.  The time axis is presented in the rest-frame and
  relative to $B$-band maximum.}
\label{f:colors}
\end{figure}

The color evolution of LSQ14fmg is also unique (Fig.~\ref{f:colors}).
Note that we present the observed colors, rather than attempting to
obtain the intrinsic colors through the very uncertain host reddening
corrections.  The observed $B-V$ color shows a completely flat
evolution and remains red ($B-V=0.14$) out to roughly one week past
$B$ maximum.  The evolution then tracks that of normal \sneia\ toward
redder colors until the epoch of the Lira Law
\citep{1996MsT..........3L, 1999AJ....118.1766P} is reached.  Our time
coverage at this epoch is insufficient to obtain a reliable $B-V$
slope.  However, the $B-V$ color tracks the blue end of the normal
\snia\ sample, perhaps indicating minimal host galaxy reddening.
Recall that E$(B-V)$ derived from \naid\ is uncertain, but is
consistent with minimal to no reddening.  The observed $r-i$ color
evolution of LSQ14fmg is even more remarkable.  A period of flat
evolution is also observed out to roughly one week past $B$ maximum
and is comparatively red ($r-i=-0.11$).  Furthermore, while $r-i$
curves of normal \sneia\ show the similar characteristic shape of the
$B-V$ curves, the $r-i$ of LSQ14fmg is completely featureless and
monotonically increasing toward redder colors with time.  Other
\superc\ objects show flatter $r-i$ curves, but never as extreme.

%NIR 

LSQ14fmg is overluminous in the NIR.  Our $YJH$ light curves do not
have adequate time coverage for any detailed analyses.  However,
taking the brightest point in each band and applying only the Milky
Way extinction correction and distance modulus, the $YJH$ peaks are as
bright as $-20.4$ mag, $-20.6$ mag, and $-20.6$ mag, respectively.  In
contrast, the $YJH$ peaks for SN~2012hr, a normal \snia\ with a
similar \dm, are $-18.3$ mag, $-18.4$ mag, and $-18.1$ mag,
respectively.  LSQ14fmg is much brighter in the NIR than any normal or
\superc\ \sneia, including SNe~2009dc \citep{2011MNRAS.412.2735T},
2012dn \citep{2016PASJ...68...68Y}, and ASASSN-15pz
\citep{2019ApJ...880...35C}.

%%%%%%%%%%%%%%%%%%
%% Spectroscopy %%
%%%%%%%%%%%%%%%%%%

\section{Spectroscopic properties}
\label{s:spec}

%intro

Three optical spectra were taken of LSQ14fmg from approximately two
weeks before to a few days past $B$ maximum.  The time series spans
more than two weeks, yet it shows very little evolution in the ions
present and their line profile shapes (Fig.~\ref{f:spec}).  The
optical spectral features of LSQ14fmg are typical of \superc\ and
normal \sneia.  Although weak, all the expected lines of IMEs are
present.  

%In particular, the presence of the \sii\ features confirms
%that LSQ14fmg is a \snia.

%comparison to super-Cs

\begin{figure}
\epsscale{1.1}
\plotone{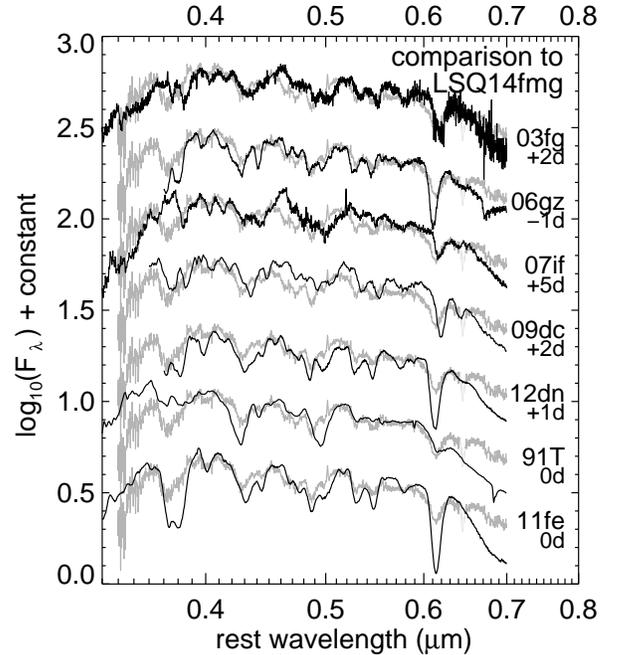}
\caption{Comparison of near-maximum-light spectra of \superc\ \sneia,
  SN~1991T, and the normal-bright SN~2011fe to that of LSQ14fmg.  The
  $+3.6$ day spectrum of LSQ14fmg is plotted in the background in gray
  for each comparison.}
\label{f:spec_compare}
\end{figure}

\begin{figure}
\epsscale{1.1}
\plotone{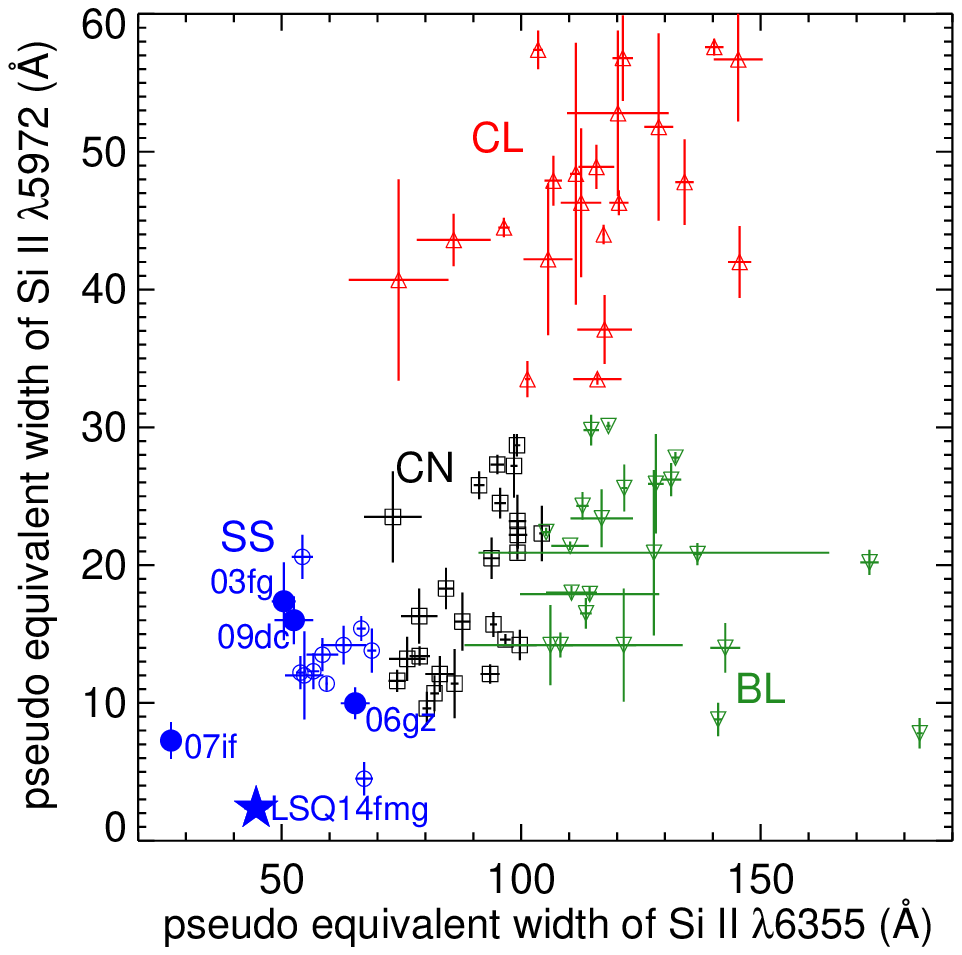}
\caption{Classification of \sneia\ based on
  \citet{2006PASP..118..560B} into ``core normal'' (CN), ``cool''
  (CL), ``broad line'' (BL), and ``shallow silicon'' (SS) subtypes.
  Measurements of the pseudo-EW were adopted from CSP-I
  \citep{2013ApJ...773...53F}.  Additional measurements of the
  \superc\ group were done using data from \citet{2006Natur.443..308H,
    2007ApJ...669L..17H, 2010ApJ...713.1073S, 2011MNRAS.412.2735T}.
  All members of the \superc\ group fall in the ``shallow silicon''
  subtype. }
\label{f:branch}
\end{figure}

In Fig.~\ref{f:spec_compare}, the near-maximum-light spectrum of
LSQ14fmg is compared to the spectra of \superc\ SNe~2003fg
\citep{2006Natur.443..308H}, 2006gz \citep{2007ApJ...669L..17H},
2007if \citep{2010ApJ...713.1073S}, 2009dc
\citep{2011MNRAS.412.2735T}, 2012dn \citep{2019MNRAS.488.5473T}, as
well as SN~1991T \citep{1992ApJ...397..304J}, and the normal-bright
SN~2011fe \citep{2014MNRAS.439.1959M}.  Comparing the spectral
features near maximum, LSQ14fmg shares many similarities with others
in the \superc\ group.  The P-Cygni profiles are much weaker compared
to those of normal \sneia, such as the well-observed, prototypical
normal \snia\ SN~2011fe in Fig.~\ref{f:spec_compare}.  The pseudo-EWs
of both \siii\ \lam\lam0.5972,0.6355~\um\ are quite small, placing
LSQ14fmg solidly in the ``shallow silicon'' group in the
\citet{2006PASP..118..560B} classification scheme, along with other
\superc\ \sneia\ (Fig.~\ref{f:branch}).  In fact, the
\siii\ \lam0.5972~\um\ line of LSQ14fmg is extremely weak in all the
spectra and was difficult to identify.  The measured velocity of the
\siii\ \lam0.6355~\um\ line is then used to locate the
\siii\ \lam0.5972~\um\ absorption.  Indeed, LSQ14fmg has the smallest
pseudo-EW of \siii\ \lam0.5972~\um\ of all the members in the
\superc\ group.

%SYNAPPS

\begin{figure}
\epsscale{1.1}
\plotone{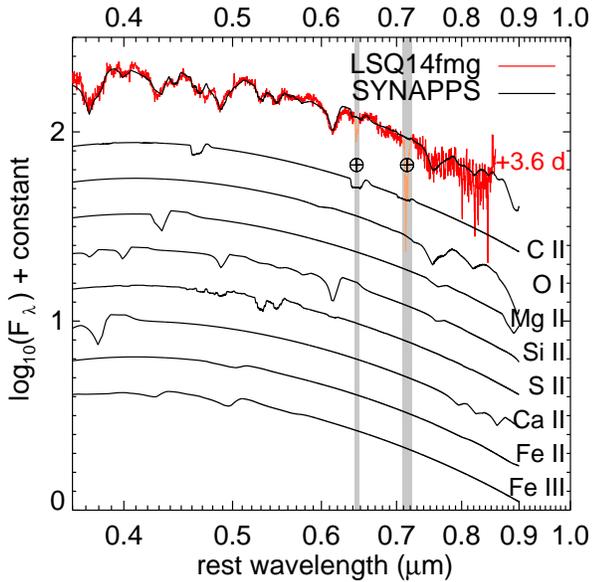}
\caption{\texttt{SYNAPPS} fit of the near-maximum spectrum of
  LSQ14fmg.  The isolated contribution of each ion is also shown.  The
  gray vertical bands mark strong telluric absorptions, which
  unfortunately coincide with \cii\ \lam\lam0.6580,0.7235~\um.}
\label{f:synapps}
\end{figure}

\texttt{SYNAPPS} \citep{2011PASP..123..237T}, a highly parameterized
and fast spectrum synthesis code derived from \texttt{SYNOW}
\citep{2005PASP..117..545B}, identified ions which are present in
normal \sneia\ in the near-maximum-light spectrum of LSQ14fmg
(Fig.~\ref{f:synapps}).  The fit shows that the spectrum is dominated
by features of IME.  Carbon may be present in the spectra of LSQ14fmg.
Unfortunately, both the \cii\ \lam\lam0.6580,0.7235~\um\ lines
coincide with telluric absorption lines.  The presence of \cii\ is
uncertain from the \texttt{SYNAPPS} fit.  Despite the shallow
\siii\ lines, LSQ14fmg does not show particularly strong
high-ionization \feiii\ lines like SN~1991T.  \superc\ \sneia, as a
group, do not appear to be at extremely high ionization states like
91T-like objects.

%Si II velocity

\begin{figure}
\epsscale{1.1}
\plotone{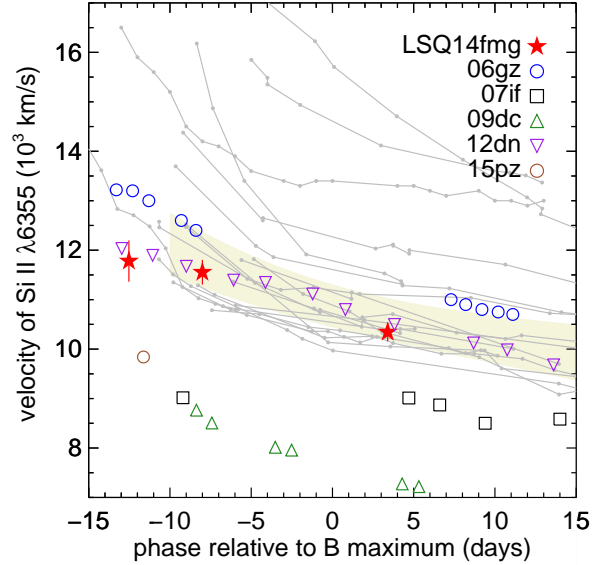}
\caption{Comparison of the time evolution of the
  \siii\ \lam0.6355~\um\ velocity between LSQ14fmg and other \sneia.
  The time axis is shown in the rest-frame of each \snia.  Normal
  \sneia\ from CSP-I are plotted as gray in the background.  The
  shaded region represents the average and the 1$\sigma$ dispersion of
  the normal sample.  \superc\ \sneia\ are highlighted.}
\label{f:vsi_t}
\end{figure}

In Fig.~\ref{f:vsi_t}, \siii\ \lam0.6355~\um\ velocities of normal and
\superc\ \sneia\ are compared.  The velocities of normal \sneia\ in
CSP-I with adequate time coverage \citep{2013ApJ...773...53F} are
plotted in the background in gray, and the velocities of SNe~2009ig
\citep{2013ApJ...777...40M} and 2011fe \citep{2013A&A...554A..27P} are
added in this group for their early-phase coverage.  The shaded region
represents the average and the 1$\sigma$ dispersion of the normal
\snia\ sample of \citet{2013ApJ...773...53F}.  Velocity measurements
of \superc\ \sneia, SNe~2006gz \citep{2007ApJ...669L..17H}, 2007if
\citep{2010ApJ...713.1073S}, 2009dc (CSP-I), 2012dn
\citep{2019MNRAS.488.5473T}, and ASASSN-15pz
\citep{2019ApJ...880...35C} are highlighted.  The \siii\ velocities of
LSQ14fmg were measured by first removing the continuum, assumed to be
a straight line connecting the blue and red boundaries of the
absorption features, then fitting a Gaussian function for the minimum.
To account for the uncertainty in determining the feature boundaries,
velocity measurements were made on $1{,}000$ realizations of varying
boundaries and random noise.  The velocity errors presented include
both the Gaussian fit error and the standard deviation of the
realizations.  Fig.~\ref{f:vsi_t} shows that there is a wide range of
\siii\ velocities for the \superc\ \sneia.  The \siii\ velocity of
LSQ14fmg is not in either extreme and is similar to those of the less
luminous objects in the \superc\ group, SNe~2006gz and 2012dn.

%Velocity plateau

The two earliest spectra of LSQ14fmg show \siii\ velocity evolving
from $11{,}800\pm400$~\kms\ to $11{,}600\pm200$~\kms, during the epoch
in which \siii\ \lam0.6355~\um\ of normal \sneia\ shows a rapid
decline in velocity (Fig.~\ref{f:vsi_t}).  The absence of the rapid
decline indicates that Si is confined at the lower-velocity range.
The velocity evolution is consistent with a monotonically decreasing
evolution similar to that of SN~2012dn \citep{2019MNRAS.488.5473T}.
From the Nearby Supernova Factory sample, \citet{2012ApJ...757...12S}
identified five overluminous \sneia\ with velocity plateaus.  These
velocity plateaus lasted at least until 10 days past $B$ maximum.
However, the ensuing drop in velocity is likely caused by the onset of
\feii\ features which blend with \siii\ \lam0.6355~\um\ and broaden
the feature toward the red, and may not represent a true decrease in
the \siii\ velocity.  The signal-to-noise ratio of our first spectrum
is too low to discern whether there is a velocity plateau.
Additionally, there is no evidence of the emergence of iron lines in
all three spectra of LSQ14fmg.  We therefore concluded that the
observed decline reflects a true \siii\ velocity decline.

%NIR spectroscopy

NIR spectroscopy can help reveal the physical properties of all types
of supernovae \citep[e.g.][]{2019ApJ...887....4D,
  2019PASP..131a4002H}.  As shown in Fig.~6 of
\citet{2019PASP..131a4002H}, NIR spectroscopy can be used to easily
identify a \superc\ \snia.  The prominent $H$-band break
\citep{2013ApJ...766...72H, 2019ApJ...875L..14A, 2019ApJ...878...86A}
is observed in normal \sneia\ starting a few days past maximum and
results from the exposed iron-group elements
\citep{1998ApJ...496..908W}.  In a \superc\ \snia, the $H$-band break
does not appear until much later, indicating that the iron-group
elements are hidden until a few months past maximum
\citep[e.g.,][]{2011MNRAS.412.2735T}.  Unfortunately, no NIR spectra
were taken of LSQ14fmg.

%%%%%%%%%%
%% Host %%
%%%%%%%%%%

\section{Host properties}
\label{s:host}

%basic properties

\begin{figure}
\epsscale{0.9} \plotone{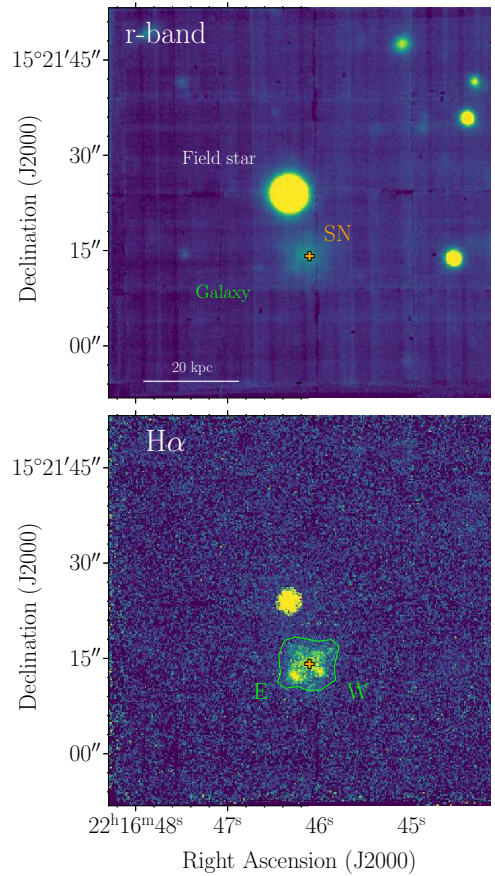}
\caption{LSQ14fmg host galaxy images extracted from the MUSE datacube.
  The top panel corresponds to the synthetic $r$-band image and the
  bottom panel the stellar-population-subtracted and
  extinction-corrected gas-phase emission line map of H$\alpha$. The
  bright central point source is a foreground field star.  The SN
  position is marked with a cross, and the host galaxy is defined by a
  green contour.}
\label{f:muse}
\end{figure}

The integral field spectroscopy obtained by MUSE was analyzed to
obtain detailed host properties.  First, a synthetic $r$-band image
was produced by convolving the transmission of the Swope $r$-band
filter with the MUSE datacube (top panel of Fig.~\ref{f:muse}).
Visual inspection of the $r$-band image at the supernova position
revealed an extended source which was presumed to be the host of
LSQ14fmg.  Based on the $r$-band brightness, a contour was defined for
the extended source (Fig.~\ref{f:muse}), and the spectrum within the
contour was then extracted from the datacube.  The spectrum shows the
typical emission lines of a star-forming galaxy on top of a blue
continuum.  By measuring the observed wavelengths of the strongest
emission lines, the heliocentric redshift for the galaxy was
determined to be $z_{\mathrm{helio}}=0.0661\pm0.0003$, in complete
agreement with the result from the WFCCD host spectrum.  The $r$-band
magnitude of the host was measured to be $r=22.40\pm0.53$ from the
datacube.  This puts the absolute magnitude of the host galaxy at
$M_r=-14.86\pm0.53$.

%STARLIGHT

The analysis of modeling the spectrum of the host galaxy of LSQ14fmg
is similar to that employed by
\citet{2014A&A...572A..38G,2016A&A...591A..48G}, using a modified
version of \texttt{STARLIGHT} \citep{2005MNRAS.358..363C,
  2016MNRAS.458..184L}.  The program models the stellar component of
the observed spectral continuum by estimating the fractional
contributions of simple stellar populations (SSP) of various ages and
metallicities, while including the effects of foreground dust.  The
base model consists of 248 spectra from the ``Granada-Miles'' (GM)
base.  GM is a combination of the MILES SSP spectra
\citep{2010MNRAS.404.1639V, 2011A&A...532A..95F} for populations older
than 63 Myr and the \citet{2005MNRAS.357..945G} models for younger
ages.  The Initial Mass Function of \citet{1955ApJ...121..161S} is
assumed in \texttt{STARLIGHT}, as well as the evolutionary tracks by
\citet{2000A&AS..141..371G}, with the exception that the youngest ages
($<$3 Myr) are based on the Geneva tracks \citep{1992A&AS...96..269S,
  1993A&AS...98..523S, 1993A&AS..101..415C}.  The GM base is defined
as a regular age, metallicity $(t,Z)$ grid with 62 ages spanning
$t=0.001-14$ Gyr and four metallicities ($Z/Z_{\odot}=$ 0.2, 0.4, 1,
and 1.5 where $Z_{\odot}=0.02$).  For the foreground dust, the
\citet{1999PASP..111...63F} reddening law with $R_V = 3.1$ is used.

%gas emission spectrum

After removing the best SSP fit from each observed spectrum, a gas
emission spectrum was obtained for each spaxel.  The flux of the most
prominent emission lines was then estimated with Gaussian fits and the
dust attenuation derived from the Balmer decrement was corrected
\citep[case B recombination of][]{2006agna.book.....O}.  From the
results at each spaxel, extinction-corrected maps of the most
prominent emission lines were created.  For example, the
extinction-corrected H$\alpha$ emission map unveils some structures on
the SE and SW parts of the galaxy (the bottom panel of
Fig.~\ref{f:muse}).  Three spectra were then extracted from the SN
location, and the SE and SW structures.  The same procedures for
fitting the spectra described above were applied to these three
spectra to obtain the properties of the SN and host environment.

%host properties measurements

The spectra of the SN location and the SE and SW structures exhibit
typical emission lines of a region with ongoing star formation, on top
of a faint stellar continuum.  This is confirmed by the location of
their line ratios in the BPT diagram \citep{1981PASP...93....5B},
falling below the \cite{2001ApJ...556..121K} demarcation.  Some
environmental properties were measured for each location and are
summarized in Table~\ref{t:muse}.  These include the ongoing
star-formation rate (SFR) as measured from the H$\alpha$ emission flux
\citep{1998ApJ...498..541K}, the weight of young-to-old populations
from the H$\alpha$ EW, and the oxygen abundance in the O3N2
\citep{2013A&A...559A.114M} and D16 \citep{2016Ap&SS.361...61D}
scales.

%host properties

Globally, a SFR of 0.081$\pm$0.027~M$_\odot$~yr$^{-1}$ and a stellar
mass of 1.15$\pm$0.26$\times$10$^9$~M$_\odot$ for the entire galaxy
were measured, corresponding to a log of the specific SFR (sSFR) of
$-10.14\pm$0.49 in yr$^{-1}$.  The gas-phase metallicities are
subsolar at all three locations considered as well as for the entire
galaxy.  The H$\alpha$ EWs are a few tens of \AA, indicating a
significant contribution of populations as young as $\sim$10 Myr
\citep[e.g.,][]{2018A&A...613A..35K}.  To place the properties of
LSQ14fmg's host galaxy in the context of other \snia\ host galaxies,
the sample of 215 host galaxies from the updated Pmas/Ppak
Integral-field of Supernova hosts COmpilation
\citep[PISCO;][]{2018ApJ...855..107G}, with properties obtained in the
same fashion, was used for comparison.  The host galaxy of LSQ14fmg
has an extremely low stellar mass (second percentile of the PISCO host
sample), similar to that of the Small Magellanic Cloud
\citep[e.g.,][]{2004ApJ...604..176S}.  The SFR is low (20th
percentile).  However, in terms of sSFR, this low-mass galaxy is very
efficient in producing new stars (90th percentile).  It contains a
relatively young stellar population component (80th percentile).  And
in both oxygen abundance calibrators, the host of LSQ14fmg ranks on
the low metallicity side (10th percentile).  These properties are
consistent with those of other host galaxies of \superc\ events
\citep[e.g.,][]{2011ApJ...733....3C}.

\begin{deluxetable}{ccccc}
\tablecolumns{5}
\tablecaption{Properties of LSQ14fmg and host environment from the 
MUSE observation.\label{t:muse}}
\tabletypesize{\scriptsize}
\tablehead{
\colhead{location} \vspace{-0.3cm}&
\colhead{H$\alpha$ EW}&
\colhead{SFR} &
\multicolumn{2}{c}{12+$\log_{10}$(O/H)} \\
&
\colhead{(\AA)} &
\colhead{(M$_\odot$ yr$^{-1}$)} &
\colhead{O3N2 (dex)} &
\colhead{D16 (dex)}
}
\startdata
LSQ14fmg & 21.94$\pm$0.45 & 0.0048$\pm$0.0018 & 8.41$\pm$0.17 & 8.30$\pm$0.12 \\
SE       & 37.31$\pm$0.64 & 0.0080$\pm$0.0016 & 8.34$\pm$0.17 & 8.20$\pm$0.12 \\
SW       & 44.11$\pm$0.69 & 0.0045$\pm$0.0017 & 8.33$\pm$0.14 & 8.22$\pm$0.13 \\
global   & 22.07$\pm$0.38 & 0.081$\pm$0.027   & 8.32$\pm$0.15 & 8.15$\pm$0.16 
\enddata
%\tablecomments{}
\end{deluxetable}

%%%%%%%%%%%%%%%%
%% Discussion %%
%%%%%%%%%%%%%%%%

\section{Discussion}
\label{s:disc}

%intro

Few theoretical studies have attempted to explain the \superc\ group,
partly because of the small sample size providing limited
observational constraints.  The peculiarities of LSQ14fmg provide
insights into their origin and have guided radiation hydrodynamical
simulations in order to determine the range of physical parameters
governing the group and the most likely explosion scenario.

\subsection{Hydrodynamical Model}

%envelope models

Models of \snia\ explosions in a dense non-degenerate envelope had
been explored by \citet{1993A&A...270..223K},
\citet{1996ApJ...457..500H}, and more recently by
\citet{2016MNRAS.463.2972N}.  They provide general explanations for
several of the key observed trends in the \superc\ group.  The
interaction between the ejecta and the dense envelope produces a
strong reverse shock.  The conversion of kinetic to luminous energy
brightens and broadens the light curve, consistent with the observed
slow rise and decline \citep{2016MNRAS.463.2972N}.  This also
decelerates the ejecta.  Thus, more massive envelopes would, in
general, manifest themselves observationally as brighter peak
magnitudes and lower photospheric velocities
\citep[e.g.,][]{2007ApJ...666.1083Q}.

%model trend

If we assume simply that the mass of the envelope is the only free
parameter governing the \superc\ subgroup, the inferred observational
trends appear to hold for the small current sample.  Within this
theoretical framework, the brightest objects (e.g., SNe~2007if and
2009dc) result from explosions within the most massive envelopes and
therefore have the lowest velocities (Fig.~\ref{f:vsi_t}), slowest
evolution (Fig.~\ref{f:LC_comp}), and the most leftover unburned
material (strongest \cii\ features) compared to the faintest objects
(e.g., SNe~2006gz and 2012dn).  These ``envelope models'' are
successful in reproducing some of the observables, but they do not
specify the physical origin of the dense envelope.  They were
originally designed to mimic the WD merger scenario.  However, these
models are also an appropriate description for the core degenerate
scenario where a WD merges with the degenerate core of an AGB star.
The envelope, in this case, would be the non-degenerate envelope of
the AGB star, rather than the accreted material from the secondary WD.

%hydrodynamic simulation

Given the above, we opted to compare the observations of LSQ14fmg to
an updated and expanded suite of envelope models of
\citet{1993A&A...270..223K} and \citet{1996ApJ...457..500H} to explore
the parameter space of the physical properties of the explosion.  The
code Hydrodynamical RAdiation (\texttt{HYDRA}) was used to simulate
the evolution toward thermonuclear runaway and the explosion itself,
as well as non-local thermodynamic equilibrium (NLTE) radiation
transport to generate model light curves and spectra \citep[][and
  references therein]{2017ApJ...846...58H}.  The amount of
\nirad\ produced depends on stellar evolution parameters such as
main-sequence mass and metallicity, as well as when the explosion
occurs \citep{2001ApJ...557..279D}.  On the other hand, the envelope
mass was treated as a free parameter, since the mechanisms for mass
loss are not well understood.  

%model for LSQ14fmg

Using the minimum \siii\ velocity as an indicator of the envelope mass
\citep{2007ApJ...666.1083Q}, the $10{,}000$~\kms\ \siii\ velocity of
LSQ14fmg (Fig.~\ref{f:vsi_t}) corresponds to an envelope mass of
roughly 0.2~\msun.  The original \texttt{DET2ENV2} model of
\citet{1996ApJ...457..500H}, i.e., 1.2~\msun\ degenerate core mass and
0.2~\msun\ non-degenerate envelope mass in pure detonation, has light
curves too faint and narrow compared to the observations of LSQ14fmg.

\begin{figure}
\epsscale{1.1} \plotone{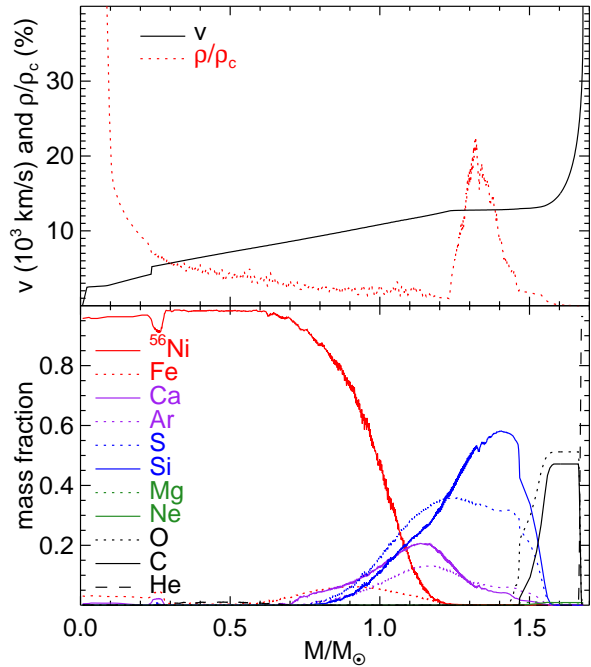}
\caption{Density and chemical structures of the model
  (\mcore~\msun\ degenerate core and \menv~\msun\ envelope) that best
  match the observations of LSQ14fmg.  The top panel plots the density
  structure relative to the initial central density, $\rho_c$, and the
  conversion between mass and velocity space.  The bottom panel plots
  the chemical structure.}
\label{f:str}
\end{figure}

Our best matching model requires a higher core mass of \mcore~\msun,
which is below the limit of $\sim2$~\msun\ for rapidly rotating WDs
\citep[e.g.,][]{2005A&A...435..967Y}, and roughly the same envelope
mass of \menv~\msun\ as \texttt{DET2ENV2}.  A higher mass of the
degenerate core is needed to produce a sufficiently slow light curve.
A star of 7~\msun\ main-sequence mass, $2\times10^{10}$~cm radius, and
metallicity Z/Z$_{\odot}=10^{-4}$ with little H and He left in the
envelope was used as a proxy for the high-mass progenitor.  The model
was also modified to include an initial deflagration burning phase
which processed \mdeflag~\msun\ of material before the ensuing
detonation \citep[e.g.,][]{2019Sci...366.7365P}.  The pre-expansion
phase is required to avoid excessive electron capture present in
classical near-Chandrasekhar-mass explosions.  A prominent shell in
the density structure is formed by the interaction with the
non-degenerate envelope within the first few minutes of the explosion
(Fig.~\ref{f:str}).  The shell then causes the confinement of
materials at lower velocities, below $12{,}000$
\kms\ (Fig.~\ref{f:str}) which matches the observations
(Fig.~\ref{f:vsi_t}).  Note that the shell is Rayleigh–Taylor unstable
which may result in mixing on a scale of approximately $2{,}000$~\kms.
The model produces \nirad\ and Si masses of \mnirad~\msun\ and
\msi~\msun, respectively, as well as small amounts of explosive
C-burning products: O, Ne, Mg (Fig.~\ref{f:str}).  In general, the
confinement at lower expansion velocities results in more efficient
gamma-ray trapping.  The results are generally high luminosities in
this class of models.

%comparison to observations

\begin{figure}
\epsscale{1.2} \plotone{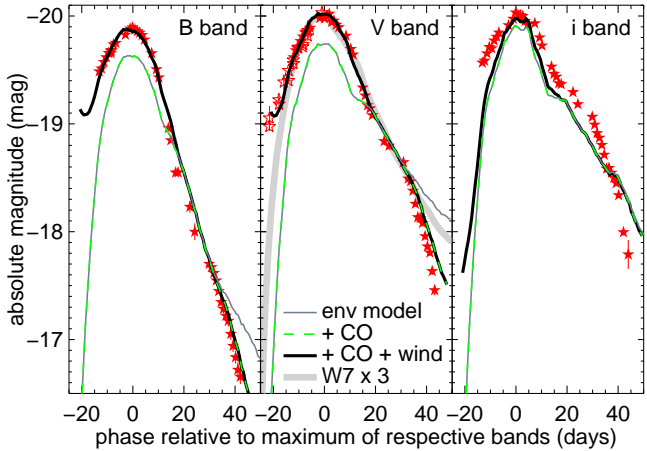}
\caption{Comparison between the $B$-, $V$-, and $i$-band light curves
  of the envelope model (\mcore~\msun\ core mass and
  \menv~\msun\ envelope mass) and LSQ14fmg (red stars).  Three model
  light curves are shown: the envelope model (thin gray curves), the
  same model with the effect of CO formation (dashed green curves),
  and with both the effects of CO formation and wind interaction
  (solid black curves).  A W7 model light curve with 3 times the mass
  is plotted as gray background to represent a \nirad\ powered light
  curve.  Note that we did not apply the uncertain host extinction
  correction for the analysis in this section and assumed that host
  extinction is negligible (Section~\ref{s:phot}).}
\label{f:model}
\end{figure}

The comparisons between the model and observed light curves are
presented in Fig.~\ref{f:model}.  The envelope model $i$-band light
curve matches the observed weak secondary maximum and the overall
light-curve shape, although the match is not perfect.  Note that the
$i$-band rise for \superc\ objects is generally slower than those in
the bluer bands (Fig.~\ref{f:LC_comp}).  The secondary maximum is
formed through the expanding atmosphere in radius, and the subsequent
decline coincides with the ionization transition
\citep{1995ApJ...444..831H, 2006ApJ...649..939K}.  For our model, the
weak secondary maximum is related to the fast receding photosphere and
photospheric conditions similar to those in subluminous \sneia.  The
envelope model $B$- and $V$-band light curves suggest an early flux
excess in LSQ14fmg that lasts until approximately one week after
maximum and a late flux deficit that starts at approximately one month
past maximum.  The model spectra show stronger NIR \ci\ lines than
optical \cii\ lines.

\subsection{Wind Interaction}

%slow rise and wind

Note that an enormous amount of \nirad\ would be required to match the
extremely slow rise of LSQ14fmg.  By doing so, it would result in
large discrepancies in the radioactive decay tail of the light curves,
\added{when the diffusion time scales are short}.  In
Fig.~\ref{f:model}, the $V$-band light curve of LSQ14fmg is compared
to a simple W7 model \citep{1984ApJ...286..644N}.  The mass of W7
needs to be increased by 3 times to approach the peak luminosity of
LSQ14fmg.  This implies a total mass of $>4$~\msun\ and pushes the IME
region to as high as $42{,}000$~\kms, which disagrees with the
observed spectra.  And even then, the rising part of the model light
curve is still not slow enough, hinting that an additional power
source is required for the early part of the observed light curve.

\begin{figure}
\epsscale{1.0} \plotone{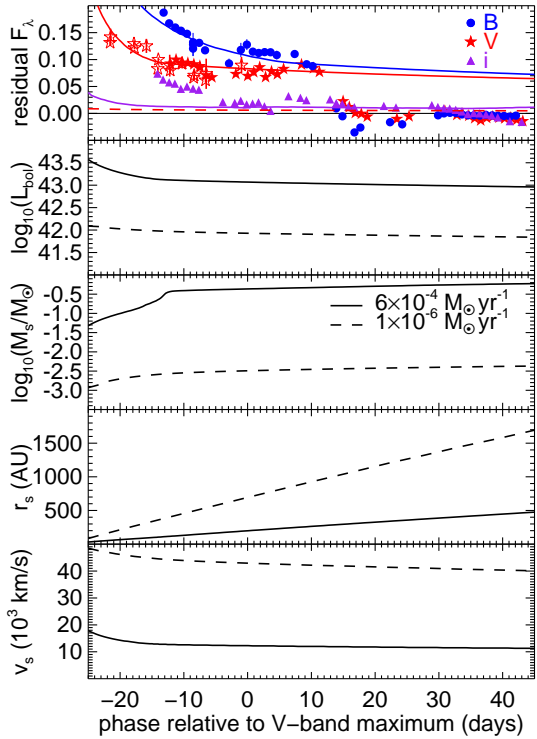}
\caption{Wind model of the excess flux observed in LSQ14fmg.  The top
  panel shows the model wind flux compared to the LSQ14fmg flux
  residual in the $B$ (blue circles), $V$ (red stars), and $i$ band
  (purple triangles) in ergs s$^{-1}$ cm$^{-2}$ \AA$^{-1}$ while
  placing the model and LSQ14fmg at 10~pc.  The open red stars
  represent LSQ $gr$ data.  The remaining four panels from top to
  bottom illustrate the shock emission bolometric luminosity,
  $L_{bol}$, in $\log_{10}$ of ergs s$^{-1}$ \AA$^{-1}$; shock-front
  mass, $M_s$, in $\log_{10}$ of \msun; shock-front radius, $r_s$, in
  AU; and velocity, $v_s$, in \kms, respectively.  The mass-loss rate
  that matches the observed excess the best
  ($6\times10^{-4}$~\msunperyr) is shown as solid curves.  The
  mass-loss rate of $1\times10^{-6}$~\msunperyr\ is shown as dashed
  curves for comparison.  Only the $V$-band flux for the
  $1\times10^{-6}$~\msunperyr\ model is shown for clarity.}
\label{f:wind}
\end{figure}

Comparing the $V$-band light curves of LSQ14fmg and our envelope model
described above (Fig.~\ref{f:model}), we found a nearly constant flux
difference in time from the earliest data point to approximately one
week past maximum (top panel of Fig.~\ref{f:wind}).  This is the same
phase and duration of the flat evolution in the observed $B-V$ and
$r-i$ color curves (Fig.~\ref{f:colors}).  The excess flux is
substantial and constitutes roughly 30\% of the modeled supernova flux
at peak in the $V$ band.  We propose that the observed flux excess is
the result of the interaction between the supernova ejecta and stellar
wind.  The spectral absorption features of LSQ14fmg are extremely weak
even when compared with other \superc\ events (Fig.~\ref{f:branch});
this may also be a result of the excess flux.  Possible reasons for
the lack of interaction signatures in the observed spectra are
explored in Section~\ref{s:cd}.

%model

We modeled the possible interaction with a stellar wind using the same
approach described in \citet{2004ApJ...607..391G} but with an updated
implementation.  The interaction efficiently converts the kinetic
energy of the rapidly expanding outer layers of the ejecta into
luminosity.  The impact velocity of the shock front, $v_{s}$, is well
in excess of $10{,}000$~\kms\ (Fig.~\ref{f:wind}), and the emission
peaks in the hard X-ray to gamma-ray regime (0.1 to 10~MeV) and then
Compton scatters.  A fraction of this hard radiation will back heat
the ejecta and deposit energy well below the optical photosphere,
e.g., at an optical depth of $5-10$.  Here, we assume that the
additional energy is thermalized, and use a bolometric correction
given by our explosion model to calculate the wind flux in the $B$,
$V$, and $i$ band.  The fact that the $B-V$ color at peak of our
supernova model matches the observed $B-V$ color of LSQ14fmg, which
contains the wind, supports the above assumption.  The transport of
hard radiation was calculated at several epochs, and we found that
roughly 1/3 of the hard radiation heats the photosphere.  The
remaining hard radiation mostly goes into expansion work.  This 1/3
factor was then used as constant in time.  Momentum conservation
($\dot{M} \times v_s = \mathrm{constant}$) dictates that the
bolometric luminosity, $L_{bol}$, starts high and declines at the
beginning (Fig.~\ref{f:wind}) as the shock-front mass, $M_{s}$, piles
up.  Thereafter, the excess emission declines slowly because $M_{s}$
now dominates the dynamics.  The time when the observed excess
emission runs out (approximately one month from explosion) is used to
estimate the outer edge of the wind, roughly $200-300$ AU in radius.
This then places constraints on the shock-front radius, $r_s$, and
$v_s$.  The wind parameters which produce the best match to the
observed flux excess are a wind speed of approximately 20~\kms\ and a
mass-loss rate of approximately $6\times10^{-4}$~\msunperyr.  This is
slightly higher than the mass-loss rate derived from the NIR excess in
SN~2012dn \citep{2017ApJ...835..143N}.

The derived mass-loss rate and wind speed are akin to those of a
typical AGB superwind \citep[e.g.,][]{2003MNRAS.341.1205V,
  2004MNRAS.355.1348M}, where the star ejects most of its mass toward
the end stage of the AGB phase \citep{1993ApJ...413..641V}.  The
duration of the excess flux is observed to last roughly one month,
which corresponds to the ejecta sweeping through $\sim10^2$ years of
wind material spanning $\sim200$ AU in radius.  This is also
consistent with the duration of a superwind episode
\citep[e.g.,][]{1997ApJ...482..897M} with a size that is much larger
than an AGB star.  If LSQ14fmg was observed in the first few days
after explosion, the wind would be optically thick and a strong early
emission could be present \citep[e.g.,][]{2016ApJ...826...96P}.  An
encounter with a superwind later in the \snia\ evolution may be
similar to the interaction signatures observed by
\citet{2019ApJ...871...62G}.  Note that any pre-existing dust in the
superwind is likely evaporated within days.

%prediction

If the superwind interpretation is correct, it is unlikely that this
is the first superwind episode of the AGB progenitor.  Assuming that
previous superwind episodes have the same mass-loss rate and duration
as derived here, we attempt to predict some observables from the
upcoming impacts here.  If the density of the medium between superwind
shells is zero due to reverse shocks, the ejecta of LSQ14fmg is
expected to produce $3-4$ rebrightening phases due to these impacts.
To estimate the X-ray and radio luminosities, Equations 51 and 52 of
\citet{2016ApJ...818...26D} were used, respectively, with the same
approximations.  As input to the equations, the time is taken as the
amount it took the superwind to reach the location of the shock
interaction and the brightness temperature is derived from the mean
kinetic energy of the ejecta at the time of impact.  To calculate the
luminosities of lower-energy photons, we assume the spectral energy
distribution of the explosion model at day 100 and consider the direct
emission from the shock material at 10~pc.  This may stand as a proxy
for the luminosity in $H_\alpha$, in the ultraviolet, or in $U$ and
$B$ bands, depending on the abundances of the ejected material.  The
results are summarized in Table~\ref{t:wind} for three upcoming
rebrightening phases and for three cases where the periods of
recurring superwind episodes are $5{,}000$, $10{,}000$, and $50{,}000$
years.  The values are reported in the rest frame of the model.  Note
that the duration of the brightening phase depends on the thickness of
the superwind shells and if the superwind shell becomes dispersed, the
luminosities given here would be correspondingly lowered.

\begin{deluxetable*}{lccccccccc}
\tablecolumns{10}
\tablecaption{Prediction of upcoming impacts with previous superwind episodes for LSQ14fmg.\label{t:wind}}
\tabletypesize{\scriptsize}
\tablehead{
\colhead{period of superwind} &
\multicolumn{3}{c}{$5{,}000$ yr} &
\multicolumn{3}{c}{$10{,}000$ yr} &
\multicolumn{3}{c}{$50{,}000$ yr}
}
\startdata
time of impact relative to explosion (yr) & 8.3 & 17 & 26 & 17 & 34 & 52 & 82 & 170 & 260 \\
X-ray luminosity ($\log_{10}$ of ergs/s) & 43.3 & 43.2 & 43.2 & 43.3 & 43.3 & 43.2 & 43.3 & 43.2 & 43.2 \\
specific radio luminosity ($10^{-19}$ ergs/s/Hz) & 1200 & 440 & 240 & 430 & 150 & 83 & 35 & 13 & 6.9 \\
specific low-energy flux ($10^{-2}$ ergs/cm$^2$/s/\AA) & 7.7 & 6.4 & 5.5 & 7.5 & 6.6 & 5.9 & 7.7 & 6.4 & 5.5 \\
\enddata
%\tablecomments{}
\end{deluxetable*}

\subsection{Formation of CO}

%CO 

\begin{figure}
\epsscale{1.0} \plotone{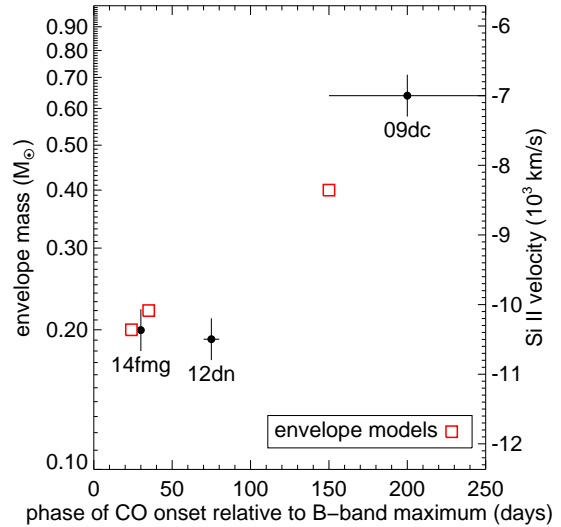}
\caption{Model prediction of the onset of CO formation in terms of
  envelope mass.  Observations are plotted using the minimum
  \siii\ velocity as an indicator of the envelope mass
  \citep{2007ApJ...666.1083Q}.  The minimum \siii\ velocities are
  taken around 5 days past maximum to avoid line blending with
  \feii\ lines at later phases.}
\label{f:co_time}
\end{figure}

Approximately $2-3$ weeks after the end of the wind interaction phase,
the $B$- and $V$-band light curves show rapid decline.  This may be
explained by active CO formation, facilitating a period of further
rapid cooling.  A \texttt{HYDRA} module for time-dependent CO
formation via neutral and charged particle reactions
\citep[e.g.,][]{1989HiA.....8..207S, 2000AJ....119.2968G} is utilized
here.  In Fig.~\ref{f:model}, the effect of CO formation is shown and
model light curves with and without CO formation begin to diverge
roughly one month past explosion in the $B$ and $V$ bands, consistent
with the observation.  As noted earlier, other \superc\ events with
adequate time coverage also show a period of rapid decline in light
curves, but at much later phases.  It is interesting to note here that
the onset of the CO formation in our envelope models is not a free
parameter.  \superc\ objects with faster-expanding ejecta, such as
LSQ14fmg, SNe~2006gz, and 2012dn, cool faster, facilitating earlier CO
formation.  On the other hand, the slower expansion rates of
SNe~2007if and 2009dc make it more difficult to cool, and the onset of
CO formation is expected to be much later.  The high mass-loss rate of
the presumed AGB progenitor of LSQ14fmg due to the recent wind episode
may have sped up the CO onset as well.  Fig.~\ref{f:co_time} shows the
model prediction of the phase of CO onset in terms of envelope mass,
and observations are plotted using the minimum \siii\ velocity as an
indicator of the envelope mass \citep{2007ApJ...666.1083Q}.  While
envelope mass is the dominant factor, note that the C/O core mass and
envelope composition also affect the phase of CO onset.  Future
follow-up efforts may use the prediction in Fig.~\ref{f:co_time} as a
rough guide to provide dense-cadence and wide-wavelength coverage at
these epochs of interest.  Note that we did not consider
\siii\ velocities later than 10 days past $B$ maximum, when iron lines
begin to emerge.

\subsection{Core-degenerate Scenario}
\label{s:cd}

%core-degenerate scenario

The observed peculiarities of LSQ14fmg all point to the progenitor
being near the final stage of AGB evolution and the core degenerate
scenario \citep[e.g.,][]{2011MNRAS.417.1466K} as the likely explosion
mechanism.  The best-matching model suggests that the progenitor has
little H and He remain in the envelope.  The early excess indicates
the interaction with a superwind, where the AGB ejects most of its
mass toward the end stage of its evolution.

%wind

The biggest weakness for the core degenerate scenario may be that no
narrow emission lines, akin to those found in Type IIn SNe
\citep{1990MNRAS.244..269S}, have been observed in a \superc\ event to
indicate the interaction between the ejecta and the AGB's envelope or
wind.  At least for LSQ14fmg, the indications are that its AGB
progenitor is near the final phase of the evolution.  Small amounts of
H and He are left in the envelope, and there may not be sufficient
optical depth in the wind.  Furthermore, gamma-ray trapping is
efficient, as is evident in the delayed onset of the $H$-band break in
the NIR spectra of \superc\ events \citep{2011MNRAS.412.2735T,
  2019PASP..131a4002H}.  Hard gamma-ray radiation from the radioactive
decay of \corad\ required to excite any remaining He may have been
prevented from escaping, suppressing the \hei\ lines
\citep[e.g.,][]{1988ApJ...335L..53G}.

%CO

The peculiar rapid decline in the light curves of LSQ14fmg
\replaced{on the presumed \corad\ radioactive decay tail}{starting at
  around one month past $B$-band maximum} indicates the onset of CO
formation.  CO formation requires a low temperature and high density
environment.  In a normal \snia, the expansion is too rapid for the
ejecta to cool down enough to form any CO.  The core degenerate
scenario may have more suitable conditions for CO formation since the
ejecta were cooled while enshrouded in a dense envelope.  Note that
the requirement for a large amount of C/O expanding at low velocities
in our model is similar to that of a subluminous \snia, which also
produces CO \citep{1995ApJ...444..831H}.  Furthermore, the copious
amount of mass loss during recent episodes may have stripped most of
the H and He layers, leaving a compact carbon-rich envelope.  Our
model then reproduces the observed early CO onset, as well as the high
$i$-band and NIR luminosities caused by the efficient redistribution
of flux toward the red by carbon.

%other evidence

In general, the low continuum polarization measured for two
\superc\ events also favors an AGB origin, since, in the WD merger
scenario, a prominent thick disk is formed along with a spherical
envelope \citep{2007MNRAS.380..933Y}.  Note, however, that
circumstellar material farther out could produce substantial continuum
polarization \citep{2018MNRAS.476.4806N}, given the asymmetric nature
of planetary nebulae.  Studies of the structures of young
\snia\ remnants, such as Kepler's supernova, also indicate interaction
with AGB wind \citep{2012ApJ...756....6P} or a planetary nebula
\citep{2020Galax...8...38C} from the progenitor system.

%trigger mechanism

Although this paper does not directly address the possible triggering
mechanism, the parameters derived from the hydrodynamical simulation
offer some clues.  Our model infers that the degenerate core mass is
high and could be above the Chandrasekhar mass but well below the
limit for a rotationally supported degenerate core.  An initial phase
of deflagration burning is required.  The deflagration phase suggests
that the degenerate core may be formed by secular accretion during a
common envelope phase \citep{2019nuco.conf..187H}, rather than a
merger on dynamical time scales, which would trigger a detonation.
The high mass suggests that the degenerate core mass is supported by
rapid rotation.  The explosion may then be triggered by the
contraction of the degenerate core through angular momentum transport.
The high progenitor mass also disfavors dynamical merger scenarios of
two massive WDs, as an accretion induced collapse is likely to result
\citep[e.g.,][]{1991ApJ...367L..19N}.

%%%%%%%%%%%%%%%%%
%% Conclusions %%
%%%%%%%%%%%%%%%%%

\section{Conclusions}
\label{s:conc}

The exaggerated light-curve properties of LSQ14fmg may help to reveal
the origin of the \superc\ group.  The light curves of LSQ14fmg rise
extremely slowly, even in comparison within the \superc\ group.  While
the most luminous \superc\ events have a typical rise time of $22-24$
days \citep{2017hsn..book..317T}, at $-23$ days relative to $B$
maximum, LSQ14fmg is already as bright as $M_V=-19.1\pm0.1$ mag, the
typical peak magnitude of a normal \snia\ (Fig.~\ref{f:wlr}).  The
light and color curves of LSQ14fmg indicate different power sources
dominating each epoch in the evolution.  The $B-V$ color stays flat
out to roughly one week past $B$ maximum, until \snia-like color
curves take over.  The comparison of the peculiar $V$-band light curve
with an envelope model yield an apparent excess flux, indicative of
interaction of the supernova ejecta with a wind-like density
structure.  The time-dependent flux excess suggests a mass-loss rate
akin to that of a typical AGB superwind.  The light curves thus have
multiple power sources from the interactions with the superwind and
the envelope, in addition to the radioactive decay of \nirad.  The
early peculiar rise is accompanied by a late rapid decline
\replaced{soon after the supernova enters the presumed
  \corad\ radioactive decay tail}{starting at around one month past
  $B$-band maximum}.  The rapid decline may be an indication of rapid
cooling by active CO formation, which is only possible with a dense
and massive C/O-rich envelope.  These observations of LSQ14fmg offer
three important clues about the \superc\ group: 1) their emission is
likely not entirely powered by \nirad; 2) the excess emission of
LSQ14fmg may point to interactions with a wind that is of an AGB
origin; 3) the conditions in the ejecta are favorable for CO
production.  All of the above point to the core degenerate scenario,
the merger of a WD and the degenerate core of an AGB, as the likely
explosion mechanism.  Our hydrodynamical simulation suggests that
LSQ14fmg explodes near the final evolutionary stage of the progenitor
AGB star.  A high degenerate core mass of \mcore~\msun\ and an initial
deflagration phase are required to match the observations.  While the
core degenerate scenario provides a coherent picture of the
observational properties of the \superc\ events as a group, there are
still questions left unanswered.  The duration of the late AGB
evolutionary phase is quite limited.  Can it account for the rate of
\superc\ events?  Is the explosion trigger associated with a violent
mass-loss event and is the preference for low-metallicity environments
associated with the mass-loss trigger?  What is the relation between
\superc, 91T-like, and 02ic-like \citep{2003Natur.424..651H} events?
Future observations of \superc\ events should take into consideration
predictions from the core degenerate scenario to support or refute it.
These observations may also provide constraints to the uncertain
mass-loss mechanisms of low-mass stars.

\acknowledgements

We thank the Las Campanas technical staff for their continued support over the years.  The CSP-II has been supported by NSF grants AST-1008343, AST-1613426, AST-1613455, and AST-1613472, as well as, the Danish Agency for Science and Technology and Innovation through a Sapere Aude Level 2 grant (PI M.S.).  
P.A.H. has been supported by NSF grant AST-1008962.  
L.G. was funded by the European Union's Horizon 2020 research and innovation programme under the Marie Sk\l{}odowska-Curie grant agreement No. 839090.  This work has been partially supported by the Spanish grant PGC2018-095317-B-C21 within the European Funds for Regional Development (FEDER).  
M.L.G. acknowledges support from the DiRAC Institute in the Department of Astronomy at the University of Washington. The DiRAC Institute is supported through generous gifts from the Charles and Lisa Simonyi Fund for Arts and Sciences, and the Washington Research Foundation.
K.K. and N.B.S. gratefully acknowledge the support of the George P. and Cynthia Woods Mitchell Institute for Fundamental Physics and Astronomy. We also thank the Mitchell Foundation for their sponsorship of the Cook's Branch Workshop on Supernovae where this science was discussed.
H.K. was funded by the Academy of Finland projects 324504 and 328898.
B.J.S. is supported by NASA grant 80NSSC19K1717 and NSF grants AST-1907570, AST-1920392, and AST-1911074.  We thank the Las Cumbres Observatory and its staff for its continuing support of the ASAS-SN project. ASAS-SN is supported by the Gordon and Betty Moore Foundation through grant GBMF5490 to the Ohio State University, and NSF grants AST-1515927 and AST-1908570. Development of ASAS-SN has been supported by NSF grant AST-0908816, the Mt. Cuba Astronomical Foundation, the Center for Cosmology and AstroParticle Physics at the Ohio State University, the Chinese Academy of Sciences South America Center forAstronomy (CAS-SACA), and the Villum Foundation.  
M.S. is supported by a project grant (8021-00170B) from the Independent Research Fund Denmark (IRFD), and by two generous grants (13261 and 28021) from VILLUM FONDEN.  Based on observations made with ESO Telescopes at the Paranal Observatory (programme 099.D-0022(A)).

\facilities{
Swope (e2v), 
du~Pont (RetroCam, WFCCD),
NOT (ALFOSC), 
VLT (MUSE),
ESO 1.0-m Schmidt (QUEST),
ASAS-SN
}

\software{
\texttt{SNooPy} \citep{2011AJ....141...19B},
\texttt{SYNAPPS} \citep{2011PASP..123..237T}
}

\end{CJK*}
\end{document}